\documentclass[a4paper]{article}
\usepackage{graphicx}
\usepackage{amssymb}
\usepackage{amsmath}
\usepackage{bbm}
\usepackage{cancel}
\usepackage{url}
\usepackage{bbold}
\usepackage{stmaryrd}

\newcommand{\mk}{{\sc Mathematica}}
\newcommand{\fr}{{\sc FeynRules}}
\newcommand{\amc}{{\sc MadGraph5\_aMC@NLO}}
\newcommand{\mg}{{\sc MadGraph5}}
\newcommand{\ml}{{\sc MadLoop}}
\newcommand{\fa}{{\sc FeynArts}}
\newcommand{\ufo}{{\sc UFO}}
\newcommand{\gosam}{{\sc GoSam}}
\def\be{\begin{equation}}
\def\ee{\end{equation}}
\def\bv{\begin{verbatim}}
\def\ev{\end{verbatim}}
\newcommand{\tos}{\shortrightarrow}
\newcommand{\bea}{\begin{eqnarray}}
\newcommand{\eea}{\end{eqnarray}}
\newcommand{\ol}{\overline}
\newcommand{\lag}{\mathcal{L}} 
\usepackage{color}
\usepackage{ulem}

\definecolor{lightgrey}{gray}{0.9}
\def\btab#1\etab{\begin{tabular}{p{45mm}p{85mm}}#1\end{tabular}}
\def\btabx#1\etabx{\begin{tabular}{p{60mm}p{50mm}}#1\end{tabular}}
\def\btaby#1\etaby{\begin{tabular}{p{15mm}p{95mm}}#1\end{tabular}}
\def\bcen{\begin{center}}
\def\ecen{\end{center}}
\def\bgfb#1\egfb{\bcen\fcolorbox{black}{lightgrey}{\parbox{138mm}{\btab#1\etab}}\ecen}
\def\bgfbx#1\egfbx{\bcen\fcolorbox{black}{lightgrey}{\parbox{138mm}{\btabx#1\etabx}}\ecen}
\def\bgfbalign#1\egfbalign{\bcen\fcolorbox{black}{lightgrey}{\parbox{138mm}{\btaby#1\etaby}}\ecen}
\begin{document}
\title{Automatic evaluation of UV and $R_2$ terms for beyond the Standard Model Lagrangians: a proof-of-principle}

\author{Celine Degrande\\
Institute for Particle Physics Phenomenology, Department of Physics\\ Durham University, Durham DH1 3LE, United Kingdom}

\date{}

\begin{flushright}
IPPP/14/58, 
DCPT/14/116
\end{flushright}

{\let\newpage\relax\maketitle}

\begin{abstract}
The computation of renormalized one-loop amplitudes in quantum field theory requires not only the knowledge of  the Lagrangian density and the corresponding Feynman rules, but also that of the ultraviolet counterterms.  More in general, and depending also on the methods used in the
actual computation of the one-loop amplitudes, additional interactions might be needed. One example is that of the $R_2$ rational terms in the OPP method. In this paper, we argue that the determination of all elements necessary for loop computations in  arbitrary models can be automated starting only from information on the  Lagrangian at the tree-level. In particular, we show how 
the $R_2$ rational and ultraviolet counterterms  for any renormalizable model can be computed with the help of a  new package, which we name  NLOCT and builds upon \fr\ and \fa. To show the potential of our approach,  we calculate all additional rules that are needed to promote a Two Higgs Doublet Model Lagrangian to one-loop computations in QCD and electroweak couplings.
\end{abstract}

\section{Introduction}

The ability of making accurate predictions for observables that can be compared with the corresponding measurements, has always been the key to test our understanding of the fundamental interactions. Perhaps even more now that the experimental confirmation of the Standard Model(SM) is more and more compelling  and the search for new phenomena widely opens. Two are the most common strategies followed. The first is the testing of predictions of the SM by  collecting precise measurements of observables possibly sensitive to new physics, an example being the campaign to determine the Higgs couplings to other SM particles. The second is the search of the effects from new particles in colliders, as now intensively done at the LHC. In this case also, the accurate knowledge of the properties, normalizations and shapes of the SM backgrounds turns out to be an essential ingredient. Furthermore, while not always needed for discovery, accurate predictions for BSM will be needed once excesses will be found and the need to clarify the actual nature of the new physics will finally arise.\\

The need for precise predictions  for hadron hadron collisions has motivated a large activity in the field of improving the techniques to compute NLO corrections, both in the QCD and electroweak sectors (see \cite{Butterworth:2014efa} for a recent review).
For many years the bottleneck of reaching NLO accuracy has been that of computing loops. Recently, several techniques have been proposed to automatically deal with loop amplitude.
A key ingredient for their evaluation is that any one-loop amplitude can be written as 
\be
A  = \sum_i d_i \text{Box}_i + \sum_i c_i \text{Triangle}_i + \sum_i b_i \text{Bubble}_i + \sum_i a_i \text{Tadpole}_i +R,
\ee 
where the Box, Triangle, Bubble and Tadpole are known scalar integrals and $R$ is the rational term. 
The OPP technique ~\cite{Ossola:2006us} has provided a simple way to compute the scalar integrals coefficients further improved by the use of multiple cuts~\cite{Britto:2004nc}. This method has been implemented in \ml~\cite{Hirschi:2011pa}  available in the \amc\ framework leading to a complete automated tool for NLO computation. So far only the SM model has been implemented despite that \ml\ is based on \mg\cite{Alwall:2011uj} for which many BSM models are available. As a matter of fact, the evaluation of the loop corrections requires two extra ingredients that so far have been added by hand in the model. The first one is the counterterms introduced by the renormalization procedure to absorb all the UV divergences arising at the one-loop level. While the divergences can be extracted from the scalar integrals, any renormalization scheme with a non-trivial finite part in the counterterms requires a careful redefinition of the fields and of the independent parameters of the model and the resolution of the renormalization conditions.  
The second missing element depends on the actual method used to perform the tensor decomposition of the loop amplitudes. In the case of OPP, it is a part of the rational term. 
In $d$ dimensions, any one-loop amplitude can be written as
\be
\ol{A}\left(\ol{q}\right) = \frac{1}{\left(2\pi\right)^4}\int d^d\ol{q}\frac{\ol{N}\left(\ol{q}\right)}{\ol{D}_0\ol{D}_1\dots\ol{D}_{m-1}},
\ee
with the propagator denominators given by $\ol{D}_i\equiv\left(\ol{q}+p_i\right)^2-m_i^2$ and where $m_i$ are the masses of the particles in the loop, $q$ is the loop momentum and $p_i$ are linear combinations of external momenta. All the quantities written with a bar live in $d$ dimensions and can therefore be split in a four dimensional part $x$ and a $d-4$ dimensional part $\tilde x$ as follow $\ol{x} \equiv x+\tilde x$.
Rational terms are finite contributions generated by the part of the integrand linear in $d-4$. One then organizes the rational part in two terms, $R_1$ and $R_2$. The rational term $R_1$ is due to the $d-4$ component of the integrand denominators and can be computed as the four-dimensional piece but using a different set of scalar integrals~\cite{Ossola:2008xq}.
The $R_2$ terms are defined as the finite part due to the $d-4$ component of the numerator
\be
R_2\equiv\lim_{\epsilon\tos0}\frac{1}{\left(2\pi\right)^4}\int d^d\ol{q}\frac{\tilde{N}\left(\tilde{q},q,\epsilon\right)}{\ol{D}_0\ol{D}_1\dots\ol{D}_{m-1}}\label{eq:r2def},
\ee
 where $\epsilon$ is defined by $d\equiv 4-2\epsilon$. We use here the 't Hooft-Veltman scheme~\cite{'tHooft:1972fi} such that all the quantities in the loop, \textit{i.e.} the loop momentum, the metric and the Dirac matrices live in $d$ dimensions:
\bea
&&\ol{\eta}^{\ol{\mu}\,\ol\nu}\ol\eta_{\ol\mu\,\ol\nu} = d\label{eq:r2me},\\
&& \ol{\gamma}^{\ol\mu} {\ol\gamma}_{\ol\mu}  = d\,\mathbb{1}\label{eq:r2ga},
\eea
where $\mathbb{1}$ is the identity matrix in Dirac space. The external momenta and polarization vectors have only four dimensional components.
The Dirac matrices in $d$ dimensions $\ol{\gamma}_{\ol{u}}$ are chosen to anti-commute with $\gamma_5$~\cite{Kreimer:1993bh,Korner:1991sx,Kreimer:1989ke}. Therefore, the cyclic property of Dirac trace has to be dropped to avoid algebraic inconsistency. The result of the evaluation of the integral in \eqref{eq:r2def} is a set of process independent Feynman rules. As a consequence, they should only be computed once for each model. The $R_2$ term are the second missing ingredient as they had to be computed so far by hand for each model. The $R_2$ terms are known for the full SM~\cite{Draggiotis:2009yb}\cite{Garzelli:2009is} and for QCD corrections to the MSSM~\cite{Shao:2012ja}. 
A package for the automatic computation of the $R_2$ terms for the SM has also been developed~\cite{Garzelli:2010fq}. \\

The purpose of this paper is to show that the procedure of determining the UV counterterms and the $R_2$ terms can be automated for any Lagrangian.  
The computation of the missing elements is done by three \mk\ packages, \fr~\cite{Alloul:2013bka}, NLOCT and \fa~\cite{Hahn:2000kx}. NLOCT is a completely new package, new functionalities have been added to \fr\ to renormalize models and output the NLO vertices in the \ufo\ format~\cite{Degrande:2011ua} while \fa\ has not been altered. The only requirement is that the model should be written in the Feynman gauge. At this stage, the package is restricted to renormalizable theories. Renormalizability is here understood strictly and not order by order like for effective field theories. Namely, the dimension of the operators in the Lagrangian should be equal to or lower than four.
Although the $R_2$ terms are not always required, the UV counterterms are needed for any one-loop computation. Therefore, the automatically generated models can be used to provide the necessary one-loop ingredients for other NLO tools than \amc\ like \gosam ~\cite{Cullen:2011ac} for example which is already using the \ufo\ format. 
As an explicit example, we consider the Two Higgs Doublet Model (2DHM). The 2HDM is a simple but important extension of the SM since it provides a well defined model to search for extra scalar particles. \\

The paper is organized as follows. The second section focuses on the renormalization of the Lagrangian and introduces the renormalization conditions for the on-shell scheme.  This scheme is easily extended to complex mass scheme to provide an appropriate treatment of the widths. The main advantage of those schemes is to avoid the evaluation of the loops on the external legs and it is used, for example, in  \ml\ to make the computation faster. The third section discusses the algorithm for the computation of the counterterms from the amplitudes. This section ends with the validation of the algorithm. The 2HDM is briefly introduced in Sect.~\ref{sec:2hdm} to fix the notation. The $R_2$ and UV counterterm vertices for the 2HDM are given in Sect.~\ref{sec:2hdmR2} and \ref{sec:2hdmUV} respectively. Finally, the conclusion is given in the last section.


\section{Renormalization}

\subsection{The renormalization constants}

In dimensional regularization UV-divergences appear as poles in $1/\epsilon$ where $d\equiv 4-2\epsilon$. In a renormalizable theory, they can absorbed by a redefinition of the free parameters and of the fields
\bea
x_0&\tos&x+\delta x \label{eq:renodef}, \nonumber\\
\phi_0&\tos&(1+\frac{1}{2}\delta Z_{\phi\phi})\phi + \sum_\chi \frac{1}{2} \delta Z_{\phi\chi} \chi,
\eea
where $x$ is an external parameter and $\phi$ and $\chi$ are fields with the same quantum numbers, the bare quantities are denoted by an additional zero subscript compared to the renormalized fields or parameters, the renormalization constant are preceded by a  $\delta$. For the fermions,  each chirality is renormalized independently. 
The external parameters are independent parameters which values should be fixed by experiments. On the contrary, internal parameters are functions of the external parameters. Internal parameters are also renormalized. However, their renormalization does not require the introduction of new renormalization constants and is fixed by their dependence on the external parameters. 
The same self renormalization constants $Z_{\phi\phi}$ are used for both the fields and their hermitian conjugates and not its conjugate as required by the complex mass scheme~\cite{Denner:2005fg}. Their imaginary parts would otherwise disappear form the hermitian Lagrangian. For example, the kinetic term of a scalar has an imaginary part if
\be
\left.\begin{array}{r}\phi_0\tos(1+\frac{1}{2}\delta Z_{\phi\phi})\phi\\
\phi_0^\dagger\tos(1+\frac{1}{2}\delta Z_{\phi\phi})\phi^\dagger\end{array}\right\}\Rightarrow \partial^\mu\phi_0\partial£_\mu\phi_0^\dagger\tos (1+\delta Z_{\phi\phi})\partial^\mu\phi\partial£_\mu\phi^\dagger
\ee
to absorb the imaginary part coming from the corresponding term of the two point loop amplitude.
On the contrary, they would be no imaginary part if the conjugated field is renormalized with the conjugate of the renormalization constant, \textit{i.e.} 
\be
\left.\begin{array}{r}\phi_0\tos(1+\frac{1}{2}\delta Z_{\phi\phi})\phi\\
\phi_0^\dagger\tos(1+\frac{1}{2}\delta Z_{\phi\phi}^*)\phi^\dagger\end{array}\right\}\Rightarrow \partial^\mu\phi_0\partial£_\mu\phi_0^\dagger\tos (1+\Re\delta Z_{\phi\phi})\partial^\mu\phi\partial£_\mu\phi^\dagger.
\ee
In the on-shell scheme, those constants are real and therefore also identical for both the fields and their conjugates. 
Similarly, external parameters in \fr\ are real and therefore renormalized by the same constants as their conjugates. Again, this is valid for both schemes even if the external parameters have complex renormalization constants as in the complex mass scheme. 
The renormalization is therefore identical for those two renormalization schemes but only the bare Lagrangian is hermitian in the complex mass scheme since the renormalization constants are complex in this scheme.
The bare Lagrangian can also be split into the renormalized one depending only on the renormalized quantities and a counterterm Lagrangian linear in the renormalization constants,
\be
\mathcal{L}_0=\mathcal{L}+\delta\mathcal{L}.
\ee
The UV counterterm vertices are extracted from the counterterm Lagrangian. The renormalization is performed into \fr\ where the Lagrangrian and all the parameters dependences are known\footnote{Except for the functions and variables related to one-loop computation, we refer to the \fr\ \cite{Alloul:2013bka} manual for a complete description of its functions and conventions.}. \\
The renormalization of the Lagrangian is performed by the function \verb+OnShellRenormalization+.  As indicated by the name, the renormalization is performed in the on-shell scheme or equivalently in the complex mass scheme. Therefore, the function renormalizes all the physical fields, all the masses and the remaining independent external parameters as in Eq.~\eqref{eq:renodef}. The function can be called once \fr\ and the model have been loaded as follows
\begin{verbatim}
Lren = OnShellRenormalization[MyLag, options]
\end{verbatim}
where \verb+MyLag+ is the Lagrangian, written in the \fr\ syntax, that needs to be renormalized.  The options for this function are listed in Table~\ref{tab:renoop}.  The output is the bare Lagrangian written as a function of the renormalized fields and parameters\footnote{All the renormalization constants are multiplied by {\tt FR\$CT}. This variable has no physical meaning but is used to keep only terms with at most one renormalization constant and to extract the counterterms in the \fa\ interface.}. Since each physical field and each mass have to be renormalized, the Lagrangian is expanded over flavor  in \fr\ before performing the renormalization procedure. \\
\begin{table}
\bgfb
\multicolumn{2}{c}{\textbf{Table~\ref{tab:renoop}: Options of \tt{OnShellRenormalization}}}\\
\\
\tt{QCDOnly} &   If {\tt{True}}, only the fields which transform under the strong gauge group, their masses and the external couplings related to the strong interaction are renormalized. The default is \tt{False}. \\
\tt{FlavorMixing}  &  If {\tt{True}}, all the fields with the same quantum numbers are allowed to mix for the renormalization. If {\tt False}, no mixing is introduced for the renormalization. If the value of this option is a list of pairs of fields ({\tt ClassMembers} should be used if present while {\tt ClassName} should be used otherwise), only the mixing between the two particles of each pair is included . The default is \tt{True}.\\
\tt{Only2Point} & If {\tt{True}}, only the fields and masses are renormalized. The default is \tt{False}.\\
\tt{Simplify2Point} & If {\tt{True}}, the mass and kinetic terms are simplified using the values of the internal parameters before renormalization is performed. The default is \tt{True}.\\
\tt{Exclude4ScalarsCT} & If {\tt{True}}, the four scalar terms are kept in the Lagrangian but not renormalized. The default is \tt{False}.
\egfb
\textcolor{white}{\caption{\label{tab:renoop}}}
\end{table}

A first observation regards the renormalization of masses that do not appear as external parameters in the \fr\ model but as internal ones. They should be exchanged for the renormalization with an external parameter appearing in its expression\footnote{the  {\tt{Value}} of its parameter definition in the \fr\ model}. This exchange can be done without changing the initial model (such that the param\_card does not change for example) using the variable \verb+FR$LoopSwitches+. The variable \verb+FR$LoopSwitches+ is a list of pairs of an external parameter and an internal parameter to be exchanged for the renormalization. Namely, the second parameter is renormalized and the first one is replaced by inverting the {\tt{Value}} of the second to obtain its dependence on the renormalization constants of the other parameters. For example, the $W$-boson mass in the \fr\ SM model~\cite{SMfr} is an internal parameter but it should be renormalized in the on-shell scheme. Consequently, one of the external parameter appearing in its expression should not have an independent renormalization constant. The command
\begin{verbatim}
FR$LoopSwitches = {{Gf, MW}};
\end{verbatim}
replaces then the Fermi constant by a function of the $W$-boson mass for the renormalization. 
Since the leading term in the expansion over a fermion mass comes from the Yukawa coupling for processes involving the Higgs boson, the \fr\  SM model defines two masses for each fermion: The Yukawa mass used to obtain the Yukawa coupling and the usual mass used anywhere else. Therefore, the mass of a fermion can be set to zero while keeping its Yukawa coupling. The Lagrangian is renormalizable and unitary only if those two external parameters are equal.  More precisely, only the masses should be used since we are using the on-shell renormalization scheme.
The variable \verb+FR$RmDblExt+ is a list of rules which is applied on the Lagrangian to replace the external parameters that should be removed before the renormalization. In the example of the SM, the command
\begin{verbatim}
FR$RmDblExt = { ymb -> MB, ymc -> MC, ymdo -> MD, yme -> Me, 
   ymm -> MMU, yms -> MS, ymt -> MT, ymtau -> MTA, ymup -> MU};
\end{verbatim}
ensures that the Lagrangian is renormalizable by replacing all the Yukawa masses by the usual ones before starting the renormalization procedure. 
The assignment of both variables should be done before running the  \verb+OnShellRenormalization+ either in the notebook or in the \fr\ model file.\\
The purpose of the options of the  function \verb+OnShellRenormalization+ listed in Table~\ref{tab:renoop} is mostly to speed up the computation by removing unwanted terms. 
\verb+QCDOnly+ keeps only the fields which carry an index of a representation of the strong gauge group, their masses and the external  parameters with  QCD appearing in their \verb+InteractionOrder+ for the renormalization, {\textit i.e.} the fields and parameters that are renormalized by the strong interaction. Similarly, the mixing between the flavors in the renormalization of the fields can be forbidden by setting {\tt FlavorMixing} to {\tt False} if the loop corrections do not mix different fields like the corrections from strong interactions in the SM for example. By default, all the fields with the same quantum numbers are allowed to mix. An intermediate situation can be obtained by setting the option {\tt FlavorMixing} to a list of pair of fields. The syntax for the fields in this list should be the \fr\ {\tt ClassMembers} if present and the {\tt ClassName} otherwise. Only the mixing between the two fields of each pair is then included. For example, only the mixing between the two light up and down quarks are included if 
\begin{verbatim}
FlavorMixing -> {{u, c}, {d, s}}.
\end{verbatim}
If three or more fields are allowed to mix, all the possible pairs of those fields should be included in the list.  
The {\tt{Only2Point}} option allows to remove the renormalization of the coupling since they are not needed afterwards if they are renormalized in the $\overline{MS}$ scheme (see Sect.~\ref{sec:solreno}).
The computation of the renormalization conditions requires that the kinetic and mass terms are diagonal for the renormalized fields. The kinetic and mass terms are therefore simplified using the values of the internal parameters. This simplification can be skip if they are already diagonal by setting the option {\tt{Simplify2Point}} to {\tt False}. Finally, the counterterms for the four scalar terms can be time consuming while they are only useful if the one of the born amplitudes contains at least one four scalar interaction. Therefore the user can remove them by setting \verb+Exclude4ScalarsCT->False+. However, the tree-level four scalar interactions are kept as they can still appear in loops.

 \subsection{Renormalization conditions}\label{sec:renocond}

While the cancelation of the divergences fixes the coefficient of the pole in a counterterm vertex, more complicated renormalization conditions may be desirable to determine its UV finite pieces.  The renormalization conditions should be chosen to ease as much as possible the problem at hand or to make the physics transparent. In this respect, the renormalized mass is identified to the physical one, the real part of the pole of the propagator in the on-shell scheme such that its value is given by the mass measurement. Furthermore it allows to get rid of the corrections on the external legs of the amplitudes by forcing the two-point functions to vanish on-shell. More details on the on-shell scheme can be found in Ref.~\cite{Denner:1991kt}.  
In the following, we will given the renormalization conditions as they are implemented in the NLOCT package. \\

First, the tadpole counterterms are chosen to cancel the loop corrections such that no tadpole should be included in any computation. Secondly, the mass and the wave functions renormalization constants are fixed by the conditions on the two-point functions.
Writing the renormalized fermion two-point function as
\be
 i\delta_{ij} \left(\cancel{p}-m_i\right) + i\left[ f^L_{ij}\left(p^2\right)\cancel{p}\gamma_-+f^R_{ij}\left(p^2\right)\cancel{p}\gamma_+ + f^{SL}_{ij}\left(p^2\right)\gamma_- + f^{SR}_{ij}\left(p^2\right)\gamma_+\right],
\ee
where $\gamma_\pm=\frac{1\pm\gamma_5}{2}$ and the $f$ functions contain both the loop and counterterm contributions, the renormalization conditions in the on-shell scheme for the fermions are
\bea
\tilde\Re\left[ f^L_{ij}\left(p^2\right) m_i + f^{SR}_{ij}\left(p^2\right)\right]\Big|_{p^2=m_i^2} &=& 0,\nonumber\\
\tilde\Re\left[ f^R_{ij}\left(p^2\right) m_i + f^{SL}_{ij}\left(p^2\right)\right]\Big|_{p^2=m_i^2} &=& 0,\nonumber\\
\tilde\Re\left[ 2 m_i\frac{\partial}{\partial p^2}\left[\left(f^L_{ii}\left(p^2\right)+f^R_{ii}\left(p^2\right)\right) m_i + f^{SL}_{ii}\left(p^2\right)+ f^{SR}_{ii}\left(p^2\right)\right] 
+ f^L_{ii}\left(p^2\right)+f^R_{ii}\left(p^2\right)\right] \Big|_{p^2=m_i^2} &=& 0.\nonumber\\
\eea
The function $\tilde\Re$ takes the real part of the loop function but not of the couplings or of the mixing parameters. The off-diagonal conditions allow to absorb the corrections that mix different flavors in the wave function renormalizations. The renormalized fields are therefore mass eigenstates. If the two fermion flavors are massless, the first two  conditions are trivially satisfied and therefore are replaced by $\tilde\Re f^L_{ij}\left(0\right)=0$ and $\tilde\Re f^R_{ij}\left(0\right)=0$ to fix the renormalization constants. For a Majorana fermions $\Psi$, the left and right renormalization constant for the wave function should be complex conjugate of each other since the left and right handed fermion fields are related by 
\be
\Psi_R = e^{i\alpha} \left(\Psi_L\right)^c
\ee 
where $\alpha$ is the Majorana phase. The two first conditions should therefore be equivalent for a Majorana fermion if only one renormalization constant is used.
Similarly,  if the renormalized two-point function for a scalar is 
\be
i \delta_{ij} \left(p^2-m_i^2\right) + i f^S_{ij}\left(p^2\right),
\ee
and the renormalization conditions read
\bea
\tilde\Re\left[ f^S_{ij} \left(p^2\right) \right]\Big|_{p^2=m_i^2}&=&0\nonumber\\
\tilde\Re\left[ f^S_{ij} \left(p^2\right) \right]\Big|_{p^2=m_j^2}&=&0\nonumber\\
\tilde\Re\left[ \frac{\partial}{\partial p^2}f^S_{ii} \left(p^2\right) \right]\Big|_{p^2=m_i^2}&=&0.
\eea
Finally, if the renormalized two-point function of a vector is written as
\be
 -i\delta_{ij} \eta_{\mu\nu} \left(p^2-m_i^2\right) - i f^T_{ij}\left(p^2\right)\left(\eta_{\mu\nu}-\frac{p_\mu p_\nu}{p^2}\right)-if^{VL}_{ij}\left(p^2\right) \frac{p_\mu p_\nu}{p^2},
\ee
the corresponding renormalization conditions are 
\bea
\tilde\Re\left[ f^T_{ij} \left(p^2\right) \right]\Big|_{p^2=m_i^2}&=&0\nonumber\\
\tilde\Re\left[ f^T_{ij} \left(p^2\right) \right]\Big|_{p^2=m_j^2}&=&0\nonumber\\
\tilde\Re\left[ \frac{\partial}{\partial p^2}f^T_{ii} \left(p^2\right) \right]\Big|_{p^2=m_i^2}&=&0.
\eea
The complex mass scheme allows the renormalized masses and the wave functions to be complex and is obtained by removing the $\tilde\Re$~\cite{Denner:2005fg}.\\

Finally, all the external parameters but the masses are renormalized in the $\ol{MS}$ scheme by default. Namely, only the pole in
\be
\frac{1}{\ol \epsilon} \equiv \frac{1}{ \epsilon} -\gamma + \log\left(4\pi\right) 
\ee
where $\gamma$ is the Euler-Mascheroni, constant is included in the counterterms. This scheme will be used for example for the Yukawa couplings to the scalar doublet without vev in the generic 2DHM. It is no longer true in a type I or II 2HDM where all the Yukawa depends on the masses in a similar way as in the SM. It will be used also for the gauge couplings or the four scalars couplings as long as they do not depend on the masses or other external parameters.
Alternatively, the zero-momentum scheme is commonly used for the renormalization of the gauge coupling constant. The renormalized coupling is fixed by requiring that the vertex between the fermions and the gauge boson is equal to the tree-level one when the momentum of the boson vanishes. 
Writing the renormalized vectorial gauge interactions of a fermion as
\begin{eqnarray}
\Gamma^\mu_{FFV}\left(p_1,p_2\right)&=&i g T^a \delta_{f_1,f_2}\Bigg[\gamma^\mu\left( \frac{\delta g}{g} + \frac{1}{2}\delta Z_{VV} + \frac{1}{2}\delta Z_{FF}^R+ \frac{1}{2}\delta Z_{FF}^L+ \frac{g'_V}{2g} \delta Z_{V'V}\right)\nonumber\\
&&+\gamma^\mu\gamma_5\left( \frac{1}{2} \delta Z_{FF}^R-\frac{1}{2} \delta Z_{FF}^L + \frac{g'_A}{2g} \delta Z_{V'V}\right)\nonumber\\
&&+ \left(\gamma^\mu h^V\left(k^2\right)+\gamma^\mu\gamma_5 h^A\left(k^2\right)+\frac{(p_1-p_2)^\mu}{2 m} h^S\left(k^2\right)+\frac{k_\mu}{2 m} h^P\left(k^2\right)\right)\Bigg], 
\end{eqnarray}
where $p_1$, $p_2$ and $k$ are the incoming momenta of the two fermions and the vector, the $h$ functions contain the loop contribution from the triangle diagrams, $g$ the gauge coupling constant and $T^a$ the generators of the gauge group and should be replaced by the charge for an abelian group. The first two terms are due to the renormalization of the tree-level vertex. The last pieces of the first two lines are due to the mixing with another vector $V'$  ($g'_V$ and $g'_A$ are its vector and axial couplings to the fermions). The renormalization conditions at zero momentum, i.e. $k=-p_1-p_2=0$ then read 
\begin{eqnarray}
\frac{\delta g}{g} + \frac{1}{2}\delta Z_{VV} + \frac{1}{2}\delta Z_{FF}^R+ \frac{1}{2}\delta Z_{FF}^L+ \frac{g'_V}{2g} \delta Z_{V'V} + h^V\left(0\right) + h^S\left(0\right) &=& 0\\
\frac{1}{2} \delta Z_{FF}^R-\frac{1}{2} \delta Z_{FF}^L + \frac{g'_A}{2g} \delta Z_{V'V}+h^A\left(0\right)&=&0.
\end{eqnarray}
Gauge invariance implies that the second is always satisfied as well as
\begin{equation}
\frac{1}{2}\delta Z_{FF}^R + \frac{1}{2}\delta Z_{FF}^L + h^V\left(0\right) + h^S\left(0\right) + \frac{g'_A}{2g} \delta Z_{V'V}=0.
\end{equation}
Consequently, the renormalization of the gauge coupling is fixed by
\be
\frac{\delta g}{g} + \frac{1}{2}\delta Z_{VV} + \frac{g'_V}{2g} \delta Z_{V'V}  + \frac{g'_A}{2g} \delta Z_{V'V} = 0.\label{eq:zeromom}
\ee
In practice, the contribution from the triangle loop diagrams should not and are not computed since they are related to the wave-functions renormalization by gauge invariance. Therefore, the zero-momentum scheme  can only be used for massless vector gauge bosons interacting with fermions.\\
In principle, other renormalization schemes can be used. In that case, however, the corresponding renormalization conditions should be implemented. In particular, renormalization schemes which only requires the computation of the two-point functions available in NLOCT like the $M_Z$ scheme~\cite{Burkhardt:1995tt,Dittmaier:2001ay,Eidelman:1995ny} can easily be included. As explained in Sect.~ \ref{sec:solreno}, the $\ol{MS}$ scheme can alternatively also be used for the two-point functions.


\section{Computation of the $R_2$ and UV counterterms}

The computation of the one-loop amplitudes is performed in NLOCT using \fa\ \cite{Hahn:2000kx} version 3.7 or above to write the amplitudes. NLOCT evaluates the terms of the amplitudes required for the computations of the $R_2$ and UV countertems and solves the renormalization conditions. Before loading those packages, the renormalized Lagrangian should be passed to \fa\ through the corresponding interface, \textit{i.e.}  
\begin{verbatim}
WriteFeynArtsOutput[Lren,options].
\end{verbatim}
The default generic file of \fa\ lorentz.gen contains optimized lorentz structure vectors and should be used whenever possible to fasten the computation. This can only be done if the option \verb+GenericFile+ has been set to \verb+False+ when calling the \verb+WriteFeynArtsOutput+ function. If this option is not set to \verb+False+, \fr\ will create a generic file that have to be used with the model file created simultaneously. Although the computation time may differ, both ways lead to the same result if the model Lorentz structures match those of lorentz.gen.
\fa\ should be loaded prior to the NLOCT package since the latter uses \fa\ to write the amplitudes. Both packages should be loaded in a new kernel as \fa\ and \fr\ cannot run on the same \mk\ kernel. They can however be used on the same \mk\ session as long as they run on two different parallel kernels. The computation of the $R_2$ and UV counterterms vertices is done by the {\tt WriteCT} function of the NLOCT package called as follows,
\begin{verbatim}
WriteCT[<model>,<genericfile>,options]
\end{verbatim}
where \verb+<model>+ and \verb+<genericfile>+ are the name of the \fa\ model files without extension (.mod and .gen) written as a string, the options are listed in Table~\ref{tab:writectop} and will be further detailed in the following. The \verb+genericfile+ should be set to lorentz if the \verb+GenericFile+ option has been set to \verb+False+ in \fr\ or to the same value as \verb+model+ otherwise. If the generic file is not given, the default value is lorentz. The \fa\ model files should be in the \fa\ model directory by default. The output file is written in the current directory and is named by the \fr\ model name with a \verb+.nlo+ extension by default. \\
\begin{table}
\bgfb
\multicolumn{2}{c}{\textbf{Table~\ref{tab:writectop}: Options of \tt{WriteCT}}}\\
\\
\tt{Output} &   The name of the output file written as a string to which the .nlo extension is added. The default is {\tt Automatic} and the \fr\ model name is used. \\
\tt{QCDOnly} &   If {\tt{True}}, only QCD corrections are kept. The default is \tt{False}. \\
\tt{Assumptions} &   A list of assumptions that will be used to simplify the expression of the UV parts. The default is an empty list. \\
\tt{ComplexMass } &    If {\tt{True}}, the complex mass scheme is used for the renormalization of the two-point functions. The default is \tt{False}. \\
\tt{MSbar} &   If {\tt{True}}, the $\ol{MS}$ scheme is used for the renormalization of the two-point functions. The default is \tt{False}. \\
\tt{ZeroMom} &   A list of pairs of an external parameter and the list of particles of the vertex which finite part should vanish to renormalize the external parameter in the zero-momentum scheme. \\
\tt{LabelInternal} &   If {\tt{True}}, the contribution of each diagram is multiplied by {\tt IPL[part]} where {\tt part} is the list of the particles appearing in the loop. Each particle only appears once in the list even its propagator has more than one occurrence in the loop. Similarly, the list of particles in the loop only contains particles and not antiparticles.  The default is \tt{True}. \\
\tt{KeptIndices} &  The list of indices that should be kept in the argument of the {\tt IPL} functions like a flavor index and unlike a color index. The default is an empty list. \\
\tt{Exclude4ScalarsCT} & If {\tt{True}}, the four scalar counterterms are not computed. The default is \tt{False}.\\
\tt{CTparameters} & If {\tt{True}}, the vertex are expressed in term of internal parameters if the size of the expressions of  the coupling constant is large. Those internal parameters and their expressions are stored in the  variable {\tt FR\$CTparam} in the output file. The default is \tt{False}.
\egfb
\textcolor{white}{\caption{\label{tab:writectop}}}
\end{table}

\subsection{Computation of the one-loop amplitudes contributions}

The algorithm starts with the evaluation of the $R_2$ parts of the one-loop amplitudes as well as the terms required for the resolution of the renormalization conditions, \textit{i.e.} the UV divergences and the UV finite contributions for amplitudes with at most two external particles. The latter will be referred to as the UV parts in the following. 
The evaluation of those terms relies on the assumptions that the model is renormalizable and written in Feynman Gauge\footnote{This gauge also used by \ml.}.
Since the maximal dimension of the operators in the Lagrangian is then four, only the vertices with one vector and two scalar or ghost fields or with three vector fields depend on momenta. Furthermore, only the fermion propagators bring additional momenta in the numerator of the integrand in the Feynman gauge. As a consequence, the power of the loop momentum in the numerator is bound by  the number of propagators. 
Only  diagrams with at most four propagators and with the sum of the dimension of the external fields lower or equal to four can diverge and therefore generate a $R_2$ and/or UV part. 
Furthermore, the amplitudes with four external particles do not have a UV divergence if a three scalars vertex or a scalar-vector-vector vertex is present. As a matter of fact, those vertices do not contain any momenta such that the highest power of the loop momentum in the numerator is lower than the number of propagator denominators. Similarly, the amplitudes with a ghost in the loop and four external scalar fields do not diverge. Therefore, all those diagrams are not generated. Although the computation can in principle be done in any gauge, the algorithm uses the feynman gauge assumption to speed up and simplify the code. The user should be aware that the result returned by the current version of NLOCT will not be correct for another gauge as diagrams would be wrongly discarded. \\

After the generation of the relevant irreducible amplitudes for a given number of external fermion, scalar and vector fields. The $R_2$ and UV parts are  computed at the generic level, \textit{i.e.} only the spin of the particles in the diagrams is specified. The vertices in \fa\ are written as
\be
\vec c\cdot \vec L = \sum_i c_i L_i,
\ee
where $\vec c$ is a vector containing the couplings and $\vec L$ is the vector of the Lorentz structures\footnote{They are stored respectively in the model and generic files.}. The Lorentz structures contain all the kinematic information: Dirac matrices, metric tensors, momenta, Levi-Civita tensors. The vector of Lorentz structures is common for all the vertices involving the same number of fermion, ghost, scalar and vector fields. On the contrary, the couplings are free from those elements but are functions of the parameters and the gauge group representations of the model specific for each vertex. 
As a result, all the elements of the amplitudes are fixed at the generic level except for the masses and those couplings. They will be replaced by their actual values later. \\

After shifting the loop momentum using Feynman parameters, the denominator becomes an even function of the loop momentum and therefore the terms of the numerator with an odd number of the loop momentum vanish after integration. Terms with less than two (four) occurrences of the loop momentum for amplitudes with three (four) propagators are also dropped as they do not induce a UV divergence.
Loop momenta are then gathered in scalar products using one of the following replacements
\bea
q^\mu q^\nu q^\rho q^\sigma &\tos& q^4\frac{1}{d(d+2)}\left(\eta^{\mu\nu}\eta^{\rho\sigma}+\eta^{\mu\rho}\eta^{\nu\sigma}+\eta^{\mu\sigma}\eta^{\rho\nu}\right),\\
q^\mu q^\nu  &\tos& q^2\frac{1}{d}\eta^{\mu\nu}.
\eea
The Dirac algebra and the contractions of the metric tensors are then performed using Eqs.~\eqref{eq:r2me} and \eqref{eq:r2ga} to obtain the two lower terms of the $\epsilon$ expansion of the numerator. 
The integration over the loop momentum generates the following $R_2$
\bea
\left.\int d^dq \frac{\epsilon }{q^2-m^2} \right|_{R_2}&=&  i \pi^2 m^2 ,\\
\left.\int d^dq \frac{\epsilon }{\left(q^2-\Delta\right)^2} \right|_{R_2}&=&  i \pi^2 ,\\
\left.\int d^dq \frac{q^2 \left(a\,\epsilon + b\right) }{\left(q^2-\Delta\right)^2} \right|_{R_2}&=&  i \pi^2 (2a-b)\Delta ,\\
\left.\int d^dq \frac{q^2\left(a\,\epsilon + b\right)}{\left(q^2-\Delta\right)^3} \right|_{R_2}&=&   i \pi^2 \left(a-\frac{1}{2}b\right) ,\\
\left.\int d^dq \frac{q^4\left(a\,\epsilon + b\right)}{\left(q^2-\Delta\right)^4} \right|_{R_2}&=&   i \pi^2  \left(a-\frac{5}{6}b\right),
\eea
where $a$ and $b$ do not depend on $\epsilon$ or the loop momentum but are polynomials of the Feynman parameters. Their UV parts are given by
\bea
\mu^{2\epsilon}\left.\int d^dq \frac{a\epsilon +b }{q^2-m^2} \right|_{UV}&=&  i \pi^2 m^2 \left(\frac{b}{\bar\epsilon}+a+b-b\log\left(\frac{m^2}{\mu^2}\right)\right) ,\\
\mu^{2\epsilon}\left.\int d^dq \frac{a\epsilon+b }{\left(q^2-\Delta\right)^2} \right|_{UV}&=&  i \pi^2 \left(a\epsilon+b\right)\left(\frac{1}{\bar\epsilon}-\log\left(\frac{\Delta}{\mu^2}\right)\right),\\
\mu^{2\epsilon}\left.\int d^dq \frac{q^2 \left(a\,\epsilon + b\right) }{\left(q^2-\Delta\right)^2} \right|_{UV}&=&  i \pi^2 \left(2a\epsilon+b\epsilon+2b\right)\left(\frac{1}{\bar\epsilon}-\log\left(\frac{\Delta}{\mu^2}\right)\right)\Delta ,\\
\mu^{2\epsilon}\left.\int d^dq \frac{q^2\left(a\,\epsilon + b\right)}{\left(q^2-\Delta\right)^3} \right|_{UV}&=&   i \pi^2 \frac{b}{\bar\epsilon}  ,\\
\mu^{2\epsilon}\left.\int d^dq \frac{q^4\left(a\,\epsilon + b\right)}{\left(q^2-\Delta\right)^4} \right|_{UV}&=&   i \pi^2 \frac{b}{\bar\epsilon},
\eea
 where the finite parts of the integrals are kept for cases with one and two propagators since only those are relevant for the one- and two-point functions. Those terms are removed if the number of external particles is bigger than two. The users can remove those terms by setting the options \verb+MSbar+ to true. Only the UV divergence are then kept and therefore all the quantities are renormalized in the $\overline{MS}$ scheme. For a single propagator, the logarithm is set to zero if the mass in its argument vanishes. Since the logarithms can induce an infrared pole for massless particle after the integration over the Feynman parameters, the terms linear in $\epsilon$ in the coefficient of the logarithm are kept when there are two propagators. The integration over the Feynman parameters is then performed for the cases of multiple propagators except for the terms with a logarithm. Those integrations are kept unevaluated until the external momentum is fixed by the renormalization conditions to handle properly 
massless particles. \\

After the computation of the required parts of the amplitudes, the masses and couplings are replaced by their values for each field insertion at the class level, \textit{i.e.} after fixing the type for each fermion, ghost, scalar and vector field in the diagrams. 
The \verb+QCDOnly+ option allows to keep only the QCD contributions in the same spirit as what was done for the renormalization. If this option is set to {\tt True}, all the diagrams without any vertex with at least three fields with a non-trivial representation under the strong gauge group are removed from the field insertion. Furthermore, only the terms with a power of $g_s$ higher than two and  higher the number of external colored particles if at least one external particle is in the adjoint representation of $SU(3)_c$.
Finally, the color algebra is performed for the triplet and octet ending the computation for the $R_2$ vertices. For any other representation like a sextet, the products of the color generators is left unevaluated.

  
\subsection{Resolution of the renormalization conditions}\label{sec:solreno}

The renormalization conditions are solved at each order in $\ol\epsilon_{UV}$ separately. The subscript $UV$ has been added to $\ol\epsilon$ to emphasize that it is only the pole due to the UV divergence. 
First, the UV divergent part of the UV counterterm vertices is simply given by the divergence of the corresponding one-loop amplitudes up to the overall sign. Writing the one loop amplitude as
\be
A^{loop} = A^{UV}\frac{1}{\ol\epsilon_{UV}} + A^{UVfin},
\ee
where $A^{UVfin}$ contains the UV finite part of the amplitude\footnote{Which may contain an IR divergence}, the UV divergent part of the corresponding counterterm vertex is $-A^{UV}\frac{1}{\ol\epsilon_{UV}} $. 
In so doing, the computation UV divergence of the renormalization constant is bypassed. The renormalization constants are therefore not used for quantities renormalized in the $\ol{MS}$ scheme as mentioned earlier. 
Secondly, the renormalization conditions of Sect.~\ref{sec:renocond} are solved for the UV finite part of the renormalized two-point functions to obtain the UV finite part of the renormalization constants $\delta X^{UVfin}$. Namely, only $A^{UVfin}$ is kept instead of the full loop amplitude in the renormalization condition.
The counterterm vertices with the renormalization constants replaced by their UV finite parts are finally added the UV divergent parts of the UV counterterm vertices to obtain the full UV counterterm vertices. For example, the gluon-gluon counterterm vertex is then written as 
\be
-i\delta^{a_1a_2} \delta Z_{gg}\left(p_1^{\mu_2}p_2^{\mu_1}-p_1\cdot p_2\eta^{\mu_1\mu_2}\right) =  -A^{UV}\frac{1}{\ol\epsilon_{UV}}-i\delta^{a_1a_2} \delta Z_{gg}^{UVfin}\left(p_1^{\mu_2}p_2^{\mu_1}-p_1\cdot p_2\eta^{\mu_1\mu_2}\right).
\ee
The tadpoles are the only exception, their counterterm vertices are directly given by the value of the full corresponding one-loop amplitudes with a minus sign since no associated renormalization constant have been introduced.
The contribution of each diagram to the UV divergence of a vertex can this way be associated with the particle in its loop. This information is kept for both the UV and $R_2$ vertices if  \verb+LabelInternal+ option is set to \verb+True+ and will be later included in the \ufo\ output by the associated \fr\ interface. 
The loop particles for the UV finite parts of the UV counterterms are however not well defined. For the UV finite parts of the UV counterterms, the particles in the loop are chosen to be those of the two-point amplitudes from which the renormalization constants have been computed. \\

The renormalized two-point functions is obtained by adding the UV finite part of the loop amplitudes and the  corresponding couterterm amplitudes, \textit{i.e.} the tree-level amplitudes with the vertices from $\delta\mathcal{L}$ . 
The integration over the Feynman parameter for the bubble logarithms is performed after writing the renormalization conditions, namely after the derivative over the external momentum has been performed if needed and after replacing the external momentum squared by the square of one of the external masses. Those logarithms can be written as 
 \be
 b_0\left(p^2,m_1,m_2\right)\equiv\int_0^1 dx \log\left(\frac{p^2(x-1)x+x\left(m_1^2-m_2^2\right)+m_2^2-i\epsilon_p}{\mu^2}\right),
 \ee
 where $m_1$ and $m_2$ are the masses of the particles in the loop, $p$ is the external momentum and $\epsilon_p$ is coming from the prescription for the propagators and is used to choose the appropriate side of the branch cut when $p^2\ge\left(m_1^2+m_2^2\right)$, \textit{i.e.} when the external particle is kinematically allowed to decay into the loop particles. This integral is nothing more than the finite part of the scalar two-point functions $B_0 = 1/\bar\epsilon -b_0$. Consequently, its expression is
 \be
  b_0\left(p^2,m_1,m_2\right) = \log\frac{m_1m_2}{\mu^2}+\frac{m_2^2-m_1^2}{p^2}\log\frac{m_2}{m_1}+\frac{m_1m_2}{p^2}\left(\frac{1}{r}-r\right)\log r -2\label{eq:b0}
 \ee
 with $r$ and $\frac{1}{r}$ being the roots of 
 \be
 x^2-\frac{m_1^2+m_2^2-p^2-i\epsilon_p}{m_1m_2}x+1=0
 \ee 
 and its derivative  expression is given by
 \be
 \frac{\partial b_0\left(p^2,m_1,m_2\right)}{\partial p^2}=\frac{m_1^2-m_2^2}{p^4}\log\frac{m_2}{m_1}+\frac{m_1m_2}{p^2}\left(\frac{1}{r}-r\right)\log r - \frac{1}{p^2}\left(1+\frac{r^2+1}{r^2-1}\log r\right).\label{eq:db0}
 \ee
 $r$ never cross the branch cut in the complex mass scheme. However, the logarithm can have a negative argument for real masses depending on the spectrum and are then replaced by
 \be
 \log\left(-x\right)\tos  \log\left(x\right) \pm i\pi,
 \ee
 where the sign of the imaginary part is fixed by $\epsilon_p$. However, the $i \pi$ term is dropped in the on-shell scheme due to the $\tilde\Re$. For $m_1=0$ and/or $p^2=0$, Eqs.~\eqref{eq:b0} and \eqref{eq:db0} seems divergent. However, the $b_0$ and its derivative expressions reduce to 
 \bea
b_0\left(p^2,0,m_2\right) &=& -2 + 2\frac{m_2^2}{p^2}\log\frac{m_2}{\mu} + \frac{p^2-m_2^2 }{p^2}\log\frac{m_2^2 - p^2}{\mu^2}\\
\frac{\partial b_0\left(p^2,0,m_2\right)}{\partial p^2}& =& \frac{m_2^2 \log \frac{m_2^2-p^2}{\mu^2}-2 m_2^2 \log \frac{m_2}{\mu}+p^2}{p^4}
 \eea
 when $m_1=0$, to
 \bea
 b_0\left(0,m_1,m_2\right)& =& -1+\frac{2 m_1^2 \log \frac{m_1}{\mu}-2 m_2^2 \log \frac{m_2}{\mu}}{m_1^2-m_2^2}\\
 \frac{\partial b_0\left(0,m_1,m_2\right)}{\partial p^2}& =&\frac{-m_1^4+m_2^4+4 m_2^2 m_1^2 \log \left(\frac{m_1}{m_2}\right)}{2 \left(m_1^2-m_2^2\right){}^3}
 \eea
 when $p^2=0$ and to
 \bea
b_0\left(0,0,m_2\right)&=& -1 + 2\log\frac{m_2}{\mu}\\
\frac{\partial b_0\left(0,0,m_2\right)}{\partial p^2}& =&\frac{-1}{2m_2^2}
 \eea
 when both vanishes.
  The $b_0$ functions only has an infrared divergence when all the arguments vanish,
  \be
 b_0\left(0,0,0\right)= \frac{1}{\bar\epsilon} 
 \ee
 while its derivative vanishes in that case. Its derivative is also IR divergent when one of the loop particle is massless and the other has the same mass as the external particle,
 \be
 \frac{\partial b_0\left(m^2,0,m\right)}{\partial p^2} = \frac{1}{2m^2}\left(\frac{1}{\bar\epsilon} + 2 - \log\frac{m^2}{\mu^2}\right).
 \ee
 Consequently, the expansion in $\epsilon$ is only done after the evaluation of those functions and just before solving the renormalization conditions to get the finite and IR divergence of the renormalization constants.
 The UV finite part is computed by default to cover all possible cases by inserting if statement in the expressions, \textit{i.e.} for all mass hierarchies and all non-zero masses are still allowed to vanish. As a result, the expressions can become quite long. The list  of assumptions passed by the \verb+Assumptions+ option is used to remove the cases that do not satisfy them. The expressions are therefore shorter and the computation is faster. The list of assumptions is written in the .nlo file and stored in the variable \verb+NLOCT$assumptions+ to remind the user that those vertices cannot be used when they are not satisfied. In the complex mass scheme, the expressions do not change depending if the external momentum is above or below the decay threshold and therefore setting \verb+ComplexMass->True+ can also reduce the size of the expressions.\\

The computation of the finite part of the gauge coupling renormalization constant is performed using Eq.~\eqref{eq:zeromom} if the   \verb+ZeroMom+ option has been used.
For example, 
\begin{verbatim}
ZeroMom -> {{aS, {F[7], V[4], -F[7]}}}
\end{verbatim}
fixes the finite part of $\alpha_S$ renormalization constant by requiring the finite part of the gluon to the up quark interaction to be zero at zero momentum. In the SM with the number of massless quarks $N_f=5$, the renormalization of the strong coupling constant in the zero-momentum scheme implies~\cite{Beenakker:2002nc}
\be
\frac{\delta\alpha_s}{\alpha_s}=\frac{\alpha_s}{2\pi}\left(\frac{N_f}{3}-\frac{11}{2}\right)\frac{1}{\ol{\epsilon}}+\frac{\alpha_s}{6\pi}\left(\frac{1}{\ol{\epsilon}}+\log\frac{\mu^2}{M_t^2}\right).
\ee


\subsection{The output}

The vertices computed by the {\tt WriteCT} function are stored in the output file as two lists of vertices in a format similar to the \fr\ one. The only difference is that \fa\ notation is kept for the color matrices since they have the advantage that the summed indices do not appear explicitly. The list with the $R_2$ terms is called \verb+R2$vertlist+ and the one with the UV counterterms \verb+UV$vertlist+. For example, the  \verb+R2$vertlist+ for the SM with only QCD corrections looks like
\begin{verbatim}
R2$vertlist = {
{{{anti[u], 1}, {u, 2}}, ((-I/12)*gs^2*
IndexDelta[Index[Colour, Ext[1]], Index[Colour, Ext[2]]]*IPL[{u, G}]*
(TensDot[SlashedP[2], ProjM][Index[Spin, Ext[1]], Index[Spin, Ext[2]]] + 
TensDot[SlashedP[2], ProjP][Index[Spin, Ext[1]], Index[Spin, Ext[2]]]))/Pi^2},
...
}
\end{verbatim}
where the dots represent the other vertices. If the option \verb+CTparameters+ has been set to \verb+True+, the \verb+FR$CTparam+ is a replacement list for the internal parameters used in \verb+UV$vertlist+. The list is empty otherwise. Additionally, all the FeynRules information about the model, the version of the NLOCT package and the date and time of generation appear commented in the header of the file. The \fa\ file names are stored into the variable \verb+CT$Model+ and \verb+CT$GenericModel+ and their definitions appear with the \verb+NLOCT$assumptions+  definition at the beginning of the output file. Finally, a variable keeps tack which interactions have been used for the loop corrections. Each interaction is associated with a value one (zero) if the $R_2$ and UV counterterm vertices (do not) contain its contributions in the list \verb+FR$InteractionOrderPerturbativeExpansion+. For example, the SM with only QCD correction has
\begin{verbatim}
FR$InteractionOrderPerturbativeExpansion = {{QCD, 1}, {QED, 0}};
\end{verbatim}  
The output file can the be loaded using the \verb+Get+ \mk\ function in a different kernel after reloading \fr\ and the model. The vertices can then be exported in a \ufo\ file using the \verb+UVCounterterms+ and \verb+R2Vertices+ option of the \verb+WriteUFO+ command, \textit{i.e.}
\begin{verbatim}
WriteUFO[MyLag, UVCounterterms -> UV$vertlist, R2Vertices -> R2$vertlist].
\end{verbatim}
The running times for the SM, MSSM \footnote{with flavor conservation} and 2HDM do not exceed a few hours on a dual core 2.4 GHz laptop with 4 Gb of RAM.

\subsection{Validation}

The validation is based mainly on the SM and MSSM for which the $R_2$ and/or the UV have been published.

\paragraph{SM (QCD):} The analytic expressions for the $R_2$ vertices due to the one-loop corrections from the strong interaction have been found in agreement with ~\cite{Draggiotis:2009yb}. The UV counterterms expressions due to the strong interaction using the on-shell scheme for the two-point functions and the zero-momentum scheme for the strong coupling constant have also been compared to \cite{Beenakker:2002nc}. It should be noted that the expressions remain the same for the complex mass scheme since at most one non-zero mass enter the computation of each wave function or mass renormalisation constant. Therefore this mass only appear in the logarithms and the branch cut is never an issue. Finally, the \ufo\ generated automatically by \fr, NLOCT and \fa\ has been used  in a recent version of \amc~\cite{Alwall:2014hca} and found in perfect agreement with the built-in version.

\paragraph{SM (EW):} The $R_2$ vertices from the electroweak corrections have been compared analytically to~\cite{Garzelli:2009is}. The electroweak corrections to the UV counterterms have been validated by comparison with~\cite{Denner:1991kt}. Again, the on-shell scheme has been used for the two-point functions while the electroweak coupling has been renormalized with the zero-momentum scheme,
\be
\frac{\delta e}{e} = \frac{1}{2}\left(\delta Z_{AA}- \frac{s_w}{c_w} \delta Z_{ZA}\right).
\ee
Many other renormalization scheme have been suggested for the electroweak coupling constant. However, only the $\ol{MS}$ and the zero-momentum are implemented so far. The electroweak interaction is responsible for most of the decay of the elementary particles and therefore the complex mass scheme can be used to handle all the widths properly. The \ufo\ with the electroweak corrections included is currently tested in \amc\ as its algorithm is currently extended to include the electroweak corrections.

\paragraph{MSSM (QCD):}
The  analytic expressions for $R_2$ vertices in the MSSM due to the strong interaction (using the \fa\ built in model to ease the comparison) from the NLOCT package have been compared to the expressions in the literature~\cite{Shao:2012ja} and found in perfect agreement. The MSSM allows to check that the Majorana fermions are handled properly. The MSSM is currently tested in \ml\ using \fr\ version of the model to be able to export it in the \ufo\ format. Contrary to the SM, various masses can be relevant for the corrections from the strong interaction to each two-point function due to the presence of the squarks and the gluino. 
%
%

\section{2HDM}\label{sec:2hdm}

The Two Higgs Doublet Model (2HDM) is one of the simplest extension of the SM. Only one scalar multiplet with the same transformation under the SM gauge symmetries as the Higgs field is added to the field content of the SM. Consequently, the 2HDM Lagrangian can be written as
\be
\lag_{2HDM} =  \left(D_\mu\Phi_1\right)^\dagger D^\mu\Phi_1 +  \left(D_\mu\Phi_2\right)^\dagger D^\mu\Phi_2 -\mathcal{V}_{2HDM}+\lag_{2HDM}^{Yuk}+\lag_{SM}^{\cancel{Higgs}} ,\label{eq:lag}
\ee
where $\Phi_1$ and $\Phi_2$ are the two scalar doublets, $\lag_{SM}^{\cancel{Higgs}} $ is the SM Lagrangian without all the terms involving the Higgs doublet, $\mathcal{V}_{2HDM}$ is the 2HDM scalar potential and $\lag_{2HDM}^{Yuk}$ contains the interactions between the fermions and the scalar fields. The most generic potential reads \cite{Branco:1999fs,Haber:2006ue,Eriksson:2009ws}
\bea
\mathcal{V}_{2HDM} &=& \mu_1 \Phi_1^\dagger \Phi_1 + \mu_2 \Phi_2^\dagger \Phi_2 + \left(\mu_3 \Phi_1^\dagger \Phi_2 + h.c.\right) + \lambda_1 \left(\Phi_1^\dagger \Phi_1\right)^2 + \lambda_2 \left(\Phi_2^\dagger \Phi_2\right)^2 \nonumber\\
&&+\lambda_3 \left(\Phi_1^\dagger \Phi_1\right) \left(\Phi_2^\dagger \Phi_2\right) + \lambda_4 \left(\Phi_1^\dagger \Phi_2\right) \left(\Phi_2^\dagger \Phi_1\right) + \left(\lambda_5 \left(\Phi_1^\dagger \Phi_2\right)^2 + h.c.\right)\nonumber\\
&&+ \Phi_1^\dagger \Phi_1\left(\lambda_6 \left(\Phi_1^\dagger \Phi_2\right) + h.c.\right) + \Phi_2^\dagger \Phi_2\left(\lambda_7 \left(\Phi_1^\dagger \Phi_2\right) + h.c.\right),\label{eq:vthdm}
\eea
where $\mu_3$, $\lambda_5$, $\lambda_6$ and $\lambda_7$ are complex while the other parameters are real. However, not all those parameters are observable since they can be modified by a change of basis,
\be
\left(\begin{array}{c}\Phi_1\\ \Phi_2\end{array}\right)\tos \left(\begin{array}{c}\Phi_1'\\ \Phi_2'\end{array}\right)  = U_\Phi\left(\begin{array}{c}\Phi_1\\ \Phi_2\end{array}\right),
\ee
where $U_\Phi$ is a two by two unitary matrix.
Contrary to the SM, the vacuum of the 2HDM does not automatically preserve $U(1)_{EM}$. This desirable property is achieved by forcing the two vacuum expectation values of the two scalar doublets to be aligned,
\be
\left\langle\Phi_1\right\rangle=\left(\begin{array}{c}0\\v_1/\sqrt2\end{array}\right) \quad \text{and} \quad \left\langle\Phi_2\right\rangle=\left(\begin{array}{c}0\\v_2/\sqrt2\end{array}\right) .
\ee
For this vacuum, the basis can be chosen such that only one of the doublets has a non vanishing vev,
\be
\left\langle\Phi_1'\right\rangle=\left(\begin{array}{c}0\\v/\sqrt2\end{array}\right) \quad \text{and} \quad \left\langle\Phi_2'\right\rangle=\left(\begin{array}{c}0\\0\end{array}\right) \quad \text{if} \quad U_\Phi= \left(\begin{array}{cc}\cos\beta & \sin\beta\\-\sin\beta & \cos\beta\end{array}\right),
\ee
where $v^2=v_1^2+v_2^2$ and $\tan\beta = v_2/v_1$. This basis is called the Higgs basis. The basis is not entirely fixed yet because the phase of $\Phi_2$ can still be changed. For example, it can be chosen such that either one parameter amongst $\mu_3$, $\lambda_5$, $\lambda_6$ and $\lambda_7$ is real or the CP transformation of the second doublet is given by 
\begin{equation}
\left(\mathcal{CP}\right) \Phi_2\left(\vec x,t\right) \left(\mathcal{CP}\right)^\dagger = \Phi_2^*\left(-\vec x,t\right).\label{eq:phi2cp}
\end{equation}

This will be our basis in the following and therefore we will drop the prime in the notation. 
The following relations have to be satisfied for this vacuum to be the minimum of the potential or equivalently to remove the tadpoles,
\bea
\mu_1 +v^2 \lambda_1&=& 0,\nonumber\\
 \Re[2\mu_3 + v^2 \lambda_6] &=&0, \nonumber\\
  \Im[2\mu_3 + v^2 \lambda_6] &=&0.
\eea
Therefore, removing the phase of $\mu_3$ or $\lambda_6$ is equivalent.
In this basis, the Yukawa Lagrangian is 
\bea
\lag^{Yuk}_{2HDM} &=& -\ol{Q}_L\cdot y^d \cdot d_R \Phi_1 - \ol{Q}_L\cdot y^u \cdot u_R \tilde\Phi_1-\ol{L}_L\cdot y^l \cdot l_R\Phi_1\nonumber\\
&& -\ol{Q}_L\cdot G^d \cdot d_R \Phi_2 - \ol{Q}_L\cdot G^u \cdot u_R \tilde\Phi_2-\ol{L}_L\cdot G^l \cdot l_R\Phi_2 + h.c.,
\eea
where $\tilde\Phi_i=\varepsilon \Phi^*$ with $\varepsilon = \left(\begin{array}{cc} 0 & 1\\-1 & 0\end{array}\right)$, $y^f$ are the Yukawa matrices like in the SM, \textit{i.e.} diagonal matrices with the diagonal entries equal to $\sqrt2\frac{m_f}{v}$ in the physical basis for the fermions while the $G^f$ matrices are free parameters. Since we have not introduced right-handed neutrinos, there are no Yukawa terms for the neutrinos.\\
The doublets have to be replaced in the Lagrangian by the component fields,
\begin{equation}
\Phi_1\equiv \left(\begin{array}{c}-iG^+\\\frac{h^n_1+i G_0+v}{\sqrt 2}\end{array}\right), \qquad \Phi_2\equiv \left(\begin{array}{c}H^+\\\frac{h^n_2+i h^n_3}{\sqrt 2}\end{array}\right),
\end{equation}
where $G_0$ and $G^+$ are the Goldstone bosons, to extract the masses.
The mass of the physical charged scalar, $H^+$, is 
\begin{equation}
m_{H^+}^2=\mu_2+\lambda_3 \frac{v^2}{2}\label{eq:mcharge}
\end{equation}
while the mass matrix of the neutral physical states reads
\begin{equation}
m_{0}^2=\left(\begin{array}{ccc}
\lambda_1 v^2&\Re{\lambda_6}\frac{v^2}{2}&-\Im{\lambda_6}\frac{v^2}{2}\\
\Re{\lambda_6}\frac{v^2}{2}&\frac{1}{4}\left(2m_{H^+}^2+\left(\lambda_4+2\Re\lambda_5\right)\right)&-\Im{\lambda_5}\frac{v^2}{2}\\
-\Im{\lambda_6}\frac{v^2}{2}&-\Im{\lambda_5}\frac{v^2}{2}&\frac{1}{4}\left(2m_{H^+}^2+\left(\lambda_4-2\Re\lambda_5\right)\right)
\end{array}\right).\label{mneutre}
\end{equation}  
The mass eigenstates $h_i$ of mass $m_{h_i}$ are obtained by the orthogonal transformation
\be
 \left(\begin{array}{c}h_1\\h_2\\h_3\end{array}\right)\equiv T\cdot\left(\begin{array}{c}h^n_1\\h^n_2\\h^n_3\end{array}\right)
\ee
with 
\be
T \equiv \left(\begin{array}{ccc}c_1&-s_1&0\\s_1&c_1&0\\0&0&1\end{array}\right)\left(\begin{array}{ccc}c_2&0&s_2\\0&1&0\\-s_2&0&c_2\end{array}\right)\left(\begin{array}{ccc}1&0&0\\0&c_3&-s_3\\0&s_3&c_3\end{array}\right),
\ee
where $c_i=\cos\theta_i$ and $s_i=\sin\theta_i$. The masses and mixing angles of the physical scalars are functions of the potential parameters such that
\be
T^T\cdot \left(\begin{array}{ccc}m_{h_1}&0&0\\0&m_{h_2}&0\\0&0&m_{h_3}\end{array}\right) \cdot T = m_0^2.\label{eq:mneq}
\ee
Equivalently, Eq.~\eqref{eq:mneq} has been solved for the parameters of the potential in the \fr\ implementation. 
As a matter of fact, the masses and mixing angles have been chosen to be external parameters to ease the on-shell renormalization. Additionally, it also ensures that the potential is positive definite. As a consequence, all the parameters of the potential except $\lambda_2$, $\lambda_3$ and $\lambda_7$ are internal parameters.
The gauge parameters are functions of the external parameters $M_Z$, $G_F$, $\alpha_S$, $\alpha_{EM}^{-1}$ as in the SM. The Yukawa matrices $y^{f}$ are internal parameters fixed by the fermions Yukawa masses\footnote{The double definition of the fermion masses is kept as in the SM for the same reason\cite{Plehn:2002vy,Berger:2003sm}.}. Finally, the real and imaginary parts of the $G^f$ coupling matrices are external parameters. 
In the type I or II 2HDM, those matrices are related to the masses of the fermions and should be renormalized accordingly. Therefore, they should be set as internal parameters depending on the masses before renormalization.\\
\begin{table}
\centering
\begin{tabular}{|l|l|l|}
\hline
Parameter &\verb+ParameterName+&Description\\
\hline
$\alpha_{EW}^{-1}$&aEWM1&Inverse of the electroweak coupling constant\\
$G_F$&Gf&Fermi constant\\
$\alpha_S$&aS&Strong coupling constant at the Z pole\\
$\lambda_2$&l2&Second quartic coupling\\
$\lambda_3$&l3&Third quartic coupling\\
$\Re\lambda_7$, $\Im\lambda_7$&l7R,l7I&Real and imaginary parts of the last quartic coupling\\
$\Re G^u$&GUR&Real part of the up Yukawa matrix\\
$\Im G^u$&GUI&Imaginary part of the up Yukawa matrix\\
$\Re G^d$&GDR&Real part of the down Yukawa matrix\\
$\Im G^d$&GDI&Imaginary part of the down Yukawa matrix\\
$\Re G^l$&GLR&Real part of the charged lepton Yukawa matrix\\
$\Im G^l$&GLI&Imaginary part of the charged lepton Yukawa matrix\\
$\theta_1$,$\theta_2$,$\theta_3$&mixh,mixh2,mixh3&Neutral scalars mixing angles\\
$M_Z$&MZ&Z mass\\
$m_d^y$, $m_s^y$, $M_b^y$&ymdo, yms, ymb&Down, strange and bottom Yukawa mass\\
$m_u^y$, $m_c^y$, $M_t^y$&ymup, ymc, ymt&Up, charm and top Yukawa mass\\
$m_e^y$, $m_\mu^y$, $m_\tau^y$&yme, ymm, ymtau&Electron, muon and tau Yukawa mass\\
$M_u$, $M_c$, $M_t$&MU, MC, MT & Up quarks masses\\
$M_d$, $M_s$, $M_b$&MD, MS, MB & Down quarks masses\\
$M_e$, $M_\mu$, $M_\tau$&Me, MMU, MTA & Charged lepton masses\\
$m_{H^+}$&mhc&Charged scalar mass\\
$m_{h_1}$, $m_{h_2}$, $m_{h_3}$& mh1, mh2, mh3& Neutral scalar masses\\
\hline
\end{tabular}
\caption{List of the external parameters of the 2HDM in the \fr\ implementation.}\label{tab:2HDMext}
\end{table}

The Lagrangian \eqref{eq:lag} is already strongly constrained unless some extra symmetries are added. For example, the constraints from flavor changing neutral currents are automatically satisfied if flavor is conserved. Flavor conservation requires all the Yukawa coupling matrices $G^f$ to  be diagonal as well as the CKM matrix.  Choosing the basis such that the CP transformation of the second doublet are given by Eq.~\eqref{eq:phi2cp}, CP conservation implies that $\mu_3$, $\lambda_5$, $\lambda_6$ and $\lambda_7$ are real and $G^f$ are hermitian. Therefore, $\theta_2$ and $\theta_3$ vanish such that only the CP-even scalars can mix. Additionally, custodial symmetry is achieved for the scalar potential either with $\lambda_4=2\lambda_5$ or with $\lambda_4=-2\lambda_5$ and $\lambda_6=\lambda_7=0$. In the first case, the masses of pseudoscalar and the charged scalar are equal while in the twisted case, the mass of one of the CP-even scalars is equal to the mass of the charged scalar~\cite{Gerard:2007kn}. 
While CP and flavor conservations can be imposed before renormalization, the custodial symmetry cannot because it is broken by the Yukawa and gauge interactions. CP and flavor conservations will be used only for the electroweak corrections to shorten the expressions.

\section{2HDM $R_2$ counterterms}\label{sec:2hdmR2}

\subsection{QCD corrections}

For the QCD corrections, we use the generic 2HDM Lagrangian without imposing flavor or CP conservation. Many of the $R_2$ vertices from the strong interactions of the 2HDM are the same as the SM one. As a matter of fact, only the vertices between the quarks and a scalar and between two gluon and one or two scalars are modified since the only new tree-level interactions involving colored particles are the quark Yukawa interactions. Only the modified or new vertices are displayed in this section. 
The  $R_2$ fermion-fermion-scalar vertices are given by,
\begin{eqnarray}
R_2^{\bar du H^-} &=&\frac{i g_s^2}{3\pi^2} \delta_{i_1 i_2}\left[-\left( {V^{CKM}}^\dagger G^u\right)_{f_1f_2}\gamma_+ + \left({G^d}^\dagger {V^{CKM}}^\dagger\right)_{f_1f_2}\gamma_-\right]\\
R_2^{\bar ud H^+} &=&\frac{i g_s^2}{3\pi^2} \delta_{i_1 i_2}\left[\left( {V^{CKM}}G^d\right)_{f_1f_2}\gamma_+-\left({G^u}^\dagger V^{CKM}\right)_{f_1f_2}\gamma_-\right]\\
R_2^{\bar u u h}&=&\frac{i g_s^2}{3\sqrt{2}\pi^2} \delta_{i_1 i_2}\Big[\left(y^u_{f_1f_2} T_{1s_3} + G^u_{f_1f_2}\left(T_{2s_3}-iT_{3s_3}\right)\right)\gamma_+\nonumber\\&&\qquad\qquad\quad+\left(y^u_{f_1f_2} T_{1s_3} + {G^u}^\dagger_{f_1f_2}\left(T_{2s_3}+iT_{3s_3}\right)\right)\gamma_-\Big]\\
R_2^{\bar d d h} &=&\frac{i g_s^2}{3\sqrt{2}\pi^2} \delta_{i_1 i_2}\Big[\left(y^d_{f_1f_2} T_{1s_3} + G^d_{f_1f_2}\left(T_{2s_3}+iT_{3s_3}\right)\right)\gamma_+\nonumber\\&&\qquad\qquad\quad+\left(y^d_{f_1f_2} T_{1s_3} + {G^d}^\dagger_{f_1f_2}\left(T_{2s_3}-iT_{3s_3}\right)\right)\gamma_-\Big],
\end{eqnarray}
where $i_k$, $f_k$ and $s_k$ are the color (in the fundamental representation), flavor and scalar indices of the $k^\text{th}$ particle in the superscript list. The scalar index distinguishes the three neutral physical scalars. If CP is conserved, the scalars $h_1$ and $h_2$ couple identically to left and right fermions while the pseudoscalar $h_3$ vertices are proportional to $\gamma_5$ as expected.\\

The $R_2$ vertices for the gluon-gluon-scalar can be written as
\begin{eqnarray}
R_2^{G G h}
&=&
-\frac{i v g_s^2}{32 \pi ^2} \delta _{a_1a_2} \eta _{\mu _1\mu _2} \Big(y_b
   \left(T_{2 s_3}-i T_{3 s_3}\right) \left(G_b\right)^*+y_t
   \left(T_{2 s_3}+i T_{3 s_3}\right) \left(G_t\right)^*\nonumber\\&&+y_b G_b
   \left(T_{2 s_3}+i T_{3 s_3}\right)+y_t
   G_t \left(T_{2 s_3}-i  T_{3 s_3}\right)+2\left( {y_b}^2 + {y_t}^2\right)
   T_{1s_3}\Big),
\end{eqnarray}
where $a_k$ is the color (in the adjoint representation) index of the $k^\text{th}$ particle and the masses of the quarks of the lightest two flavors have been put to zero. \\

Finally, the $R_2$ vertices between two gluons and two scalars are
\begin{eqnarray}
R_2^{ G G H^-  G^+}
&=&-\frac{g_s^2}{16 \pi ^2} \delta _{{a}_1{a}_2} \eta _{\mu _1\mu _2} V^{CKM}_{3,3}  {V^{CKM}_{3,3}}^\dagger \left(y_b  \left(G_b\right)^\dagger+y_t G_t\right)\\
R_2^{GG H^+  G^-}
&=&\frac{g_s^2}{16 \pi ^2} \delta _{{a}_1{a}_2} \eta _{\mu _1\mu _2} V^{CKM}_{3,3}  {V^{CKM}_{3,3}}^\dagger \left(y_b  G_b+y_t \left(G_t\right)^\dagger\right)\\
R_2^{ GG H^-  H^+}
&=&-\frac{i g_s^2}{16 \pi ^2} \delta _{{a}_1{a}_2} \eta _{\mu _1\mu _2} \text{tr}\left({G^d}^\dagger G^d + {G^u}^\dagger G^u\right)
\end{eqnarray}
\begin{eqnarray}
R_2^{ GG G_0  h}
 &=&-\frac{g_s^2}{32 \pi ^2} \delta _{a_1 a_2} \eta _{\mu _1\mu _2} \Big(y_b G_b \left(T_{2s_4}+i T_{3s_4}\right)-y_t G_t \left(T_{2s_4}-i T_{3s_4}\right) \nonumber\\
   &&+y_t
   \left(T_{2s_4}+i T_{3s_4}\right) {G_t}^\dagger-y_b \left(T_{2s_4}-i T_{3s_4}\right) {G_b}^\dagger\Big)\\
R_2^{ G G h  h}
   &=&-\frac{i g_s^2}{32 \pi ^2} \delta _{a_1a_2} \eta _{\mu _1\mu _2}\Bigg[2 \text{tr}\left(G^u {G^u}^\dagger+G^d {G^d}^\dagger\right)   \left(T_{2s_3} T_{2s_4}+T_{3s_3} T_{3s_4}\right)\nonumber\\
   &&+\left(G_t{}^\dagger y_t + G_b y_b\right) \left(\left(T_{2s_3}+i T_{3s_3}\right) T_{1s_4}+T_{1s_3} \left(T_{2s_4}+i T_{3s_4}\right)\right)\nonumber\\
   &&+\left(G_t y_t + {G_b}^\dagger y_b\right) \left(\left(T_{2s_3}-i T_{3s_3}\right) T_{1s_4}+T_{1s_3} \left(T_{2s_4}-i T_{3s_4}\right)\right)\nonumber\\
   &&+2 \left(\left(y_t\right)^2+\left(y_b\right)^2\right) T_{1s_3} T_{1s_4} 
   \Bigg].
 \end{eqnarray}  
 The vertices $GG h_3$, $GG G_0h_{1/2}$ and $GG h_3h_{1/2}$ vanish when CP is conserved as it should since they involve only one pseudoscalar.
 
 \subsection{EW corrections}
 
 The electroweak corrections are much longer than the QCD ones. Therefore, CP and flavor conservation are assumed for the EW corrections. The fermion and vector two-point $R_2$ vertices are not modified compared to the SM. The two-point scalar $R_2$ vertices are given by 
 \bea
R_2^{G_0 G_0}&=&\frac{i }{768 \pi ^2 c_w^2 s_w^2}
 \left[-48 c_w^2 s_w^2 \left(p^2 \left(y_b{}^2+y_t{}^2\right)-3 v^2  \left(y_b{}^4+y_t{}^4\right)\right)\right.\nonumber\\
   &&\left.+e^2 \left(3 m_{h_1}^2+3 m_{h_2}^2+3 c_{21}  \left(m_{h_1}^2-m_{h_2}^2\right)-24 M_W^2 c_w^2-18  M_Z^2+\left(4 s_w^2-6\right) p^2\right)\right]\qquad\\
 R_2^{ G^- G^+} &=&\frac{i}{768 \pi ^2 c_w^2 s_w^2}
 \left(3 c_w^2 \left(8 s_w^2 \left(y_b{}^2+y_t{}^2\right) \left(3 v^2  \left(y_b{}^2+y_t{}^2\right)-2 p^2\right)\right.\right.\nonumber\\
   &&\left.\left.+e^2 \left(m_{h_1}^2+m_{h_2}^2+ c_{21}     \left(m_{h_1}^2-m_{h_2}^2\right)-2  \left(M_Z^2 \left(2-9 s_w^2\right)+5 M_W^2+p^2\right)\right)\right)\right.\nonumber\\
   &&\left.-2 e^2 s_w^2 \left(9 M_Z^2 s_w^2+45   M_W^2+p^2\right)\right)    \\
R_2^{ H^- G^+} &=&
   \frac{ 4 \left(y_b G_b+y_t G_t\right) \left(3 v^2  \left({y_b}^2+{y_t}^2\right)-2 p^2\right)- \frac{e^2c_1 s_1}{s_w^2}   \left(m_{h_1}^2-m_{h_2}^2\right)}{128 \pi ^2}\\
  R_2^{G^- H^+} &=&
   -\frac{ 4 \left(y_b G_b+y_t G_t\right) \left(3 v^2  \left({y_b}^2+{y_t}^2\right)-2 p^2\right)- \frac{e^2c_1 s_1}{s_w^2}   \left(m_{h_1}^2-m_{h_2}^2\right)}{128 \pi ^2}\\
  R_2^{ H^- H^+} &=&
   -\frac{i}{384 \pi ^2} \left(8p^2\left(3 R^d +  R^l + 3 R^u\right) +e^2 p^2\left(\frac{ s_w^2}{c_w^2}+\frac{3}{s_w^2}+1\right)\right.\nonumber\\
   &&- 36 v^2 \left({y_b}^2+{y_t}^2\right) \left( {G_t}^2 +  {G_b}^2\right)
   -3 e^2 m_{H^+}^2\left( \frac{s_w^2  }{c_w^2}+\frac{1}{s_w^2}+1\right)\nonumber\\
   &&\left. -\frac{3  e^2}{s_w^2}\left(s_1^2 m_{h_1}^2+c_1^2   m_{h_2}^2+m_{h_3}^2\right)
   +\frac{18 e^2 M_W^2}{s_w^2}
   +9 e^2 M_Z^2\left(\frac{   s_w^2}{c_w^2}+\frac{1}{s_w^2}-3\right)   \right)\\
 R_2^{ G_0 h_2}&=&
 \frac{-i\left(  y_b   G_b\left(p^2-3 v^2   {y_b}^2\right)+ y_t   G_t\left(p^2-3 v^2 {y_t}^2\right) +\frac{e^2 c_1 s_1}{8c_w^2 s_w^2}   \left(m_{h_1}^2-m_{h_2}^2\right)\right)}{16 \pi ^2 } \\
R_2^{ h_1h_1}&=&-\frac{i}{48 \pi ^2} \bigg[s_1^2 \left(p^2 \left(3 R^d+R^l+3 R^u\right)-9 v^2 \left({y_b}^2
   {G_b}^2+{y_t}^2 {G_t}^2\right)\right)\nonumber\\
   &&-3   s_{21} \left(y_b G_b \left(p^2-3 v^2 {y_b}^2\right)+y_tG_t \left(p^2-3 v^2 {y_t}^2\right)\right)\nonumber\\
   &&+3 c_1^2 \left(p^2   \left(y_b{}^2+y_t{}^2\right)-3 v^2   \left(y_b{}^4+y_t{}^4\right)\right)\bigg]\nonumber\\
  && +\frac{i e^2 \left(s_1^2 \left(6 c_w^2 m_{H^+}^2+3 m_{h_3}^2\right)+ p^2 \left(2
   s_w^2-3\right)-3
   \left(7 c_{1}^2+3\right) M_Z^2 \left(2 c_w^4+1\right)\right)}{384 \pi ^2 c_w^2 s_w^2}\\
R_2^{h_1 h_2}&=&\frac{i}{96 \pi ^2} \bigg[6 c_{21}
   \left(y_b G_b \left(3 v^2 {y_b}^2-p^2\right)+y_t G_t
   \left(3 v^2 {y_t}^2-p^2\right)\right)\nonumber\\
 && s_{21} \left(-3 {y_t}^2
   \left(3 v^2 {G_t}^2+p^2\right)-3 {y_b}^2 \left(3 v^2{G_b}^2 + p^2 \right)  +9 v^2\left( {y_t}^4+{y_b}^4\right)\right.\nonumber\\
   &&\left.+3 p^2 R^d+p^2 R^l+3 p^2 R^u\right)\bigg]-\frac{i e^2 s_{21} \left(2 c_w^2 m_{H^+}^2+m_{h_3}^2+7 M_Z^2 \left(2 c_w^4+1\right)\right)}{256 \pi ^2 c_w^2 s_w^2}\\
R_2^{ h_2h_2}&=&
   -\frac{i}{96 \pi ^2} \bigg[2 c_1^2 \left(p^2 \left(3 R^d+R^l+3 R^u\right)-9 v^2 {y_t}^2 {G_t}^2-9 v^2 {y_b}^2 {G_b}^2\right)\\
   &&+3 \left({y_b}^2 \left(2s_{1}^2 p^2-6 v^2  s_1^2y_b{}^2\right)+  y_t^2\left(2   p^2 s_1^2 -6 s_1^2 v^2 y_t^2\right)\right.\\
  &&\left. +2 s_{21} \left(y_b G_b \left(p^2-3 v^2
   y_b{}^2\right)+y_t G_t \left(p^2-3 v^2 y_t{}^2\right)\right)\right)\bigg]\nonumber\\
   &&
   \frac{i e^2 \left(3 \left(c_{21}+1\right) \left(m_{h_3}^2+2c_w^2 m_{H^+}^2\right)+3
   \left(7 c_{21}-13\right) \left(2 c_w^4+1\right) M_Z^2-2 p^2 \left(2
   c_w^2+1\right)\right)}{768 \pi ^2 c_w^2 s_w^2}\\
R_2^{ h_3 h_3}&=&
   \frac{i e^2}{768 \pi ^2 c_w^2 s_w^2} \left(2 \left(6 c_w^2 m_{H^+}^2-9 M_Z^2 \left(2 c_w^4+1\right)+p^2 \left(2s_w^2-3\right)\right)-3 \left(c_{21}-1\right) m_{h_1}^2\right.\nonumber\\
   &&\left.+3 \left(c_{21}+1\right)m_{h_2}^2\right)
   -\frac{i \left(p^2 \left(3 R^d+R^l+3 R^u\right)-9 v^2 \left(G_b^2 y_b^2+G_t^2   y_t^2\right)\right)}{48 \pi ^2},
\eea
 where $c_{21}=\cos2\theta_1$, $R^f =  {G_{1,1}^f}^2+{G_{2,2}^f}^2+{G_{3,3}^f}^2$, $y_t=y^u_{3,3}$, $y_b=y^d_{3,3}$, $G_t=G^u_{3,3}$, $G_b=G^d_{3,3}$. \\
 
 The $R_2$ vertices for the interactions between the fermions and one scalar are given by 
 \bea
R_2^{\bar bb G_0}&=& -\frac{\delta _{i_1 i_2} \gamma ^5 \left(36 c_w^2 s_w^2 y_t \left(G_b G_t+y_b y_t\right)+e^2 y_b \left(27-34 s_w^2\right)\right)}{1152
   \sqrt{2} \pi ^2 c_w^2 s_w^2}\\
R_2^{\bar t t G_0}&=&\frac{\delta _{i_1 i_2} \gamma _{\text{s}_1,\text{s}_2}^5 \left(y_t \left(e^2 \left(14
   s_w^2+27\right)-36 s_w^2 c_w^2  y_b^2\right)-36 G_b
   y_b G_t s_w^2 \left(s_w^2-1\right)\right)}{1152 \sqrt{2} \pi ^2 c_w^2 s_w^2}\\
R_2^{\bar u u h_1}&=&  -\frac{i \delta _{i_1 i_2} }{1152 \sqrt{2} \pi ^2 c_w^2 s_w^2}
 \left(s_1 \left(36 c_w^2 G_d s_w^2   \left(G_d G_u+y_d y_u\right)+e^2 G_u \left(14 s_w^2+27\right)\right)\right.\nonumber\\
 &&\left.+c_1 \left(2 s_w^2   \left(-18 c_w^2 y_d \left(G_d G_u+y_d y_u\right)-7 e^2 y_u\right)-27 e^2   y_u\right)\right)\\
R_2^{\bar d d h_1}&=&  \frac{i \delta _{i_1 i_2}}{1152   \sqrt{2} \pi ^2 c_w^2 s_w^2}
\left(s_1 \left(e^2 G_d \left(34   s_w^2-27\right)-36 c_w^2 G_u s_w^2 \left(G_d G_u+y_d y_u\right)\right)\right.\nonumber\\
&&\left.+c_1 \left(36 c_w^2   s_w^2 y_u \left(G_d G_u+y_d y_u\right)+e^2 y_d \left(27-34 s_w^2\right)\right)\right)\\
R_2^{\bar l l h_1}&=&-\frac{i e^2 s_1 \left(14 s_w^2+3\right) G_l }{128
   \sqrt{2} \pi ^2 c_w^2 s_w^2}  \\
R_2^{\bar u u h_3}&=&  \frac{\delta _{i_1 i_2} \gamma^5 \left(36 c_w^2 G_d s_w^2 \left(G_d   G_u+y_d y_u\right)+e^2 G_u \left(14 s_w^2+27\right)\right)}{1152 \sqrt{2} \pi ^2 c_w^2   s_w^2}\\
R_2^{\bar d d h_3}&=&  -\frac{\delta _{i_1 i_2} \gamma^5 \left(36 c_w^2 G_u s_w^2 \left(G_d
   G_u+y_d y_u\right)+e^2 G_d \left(27-34 s_w^2\right)\right)}{1152 \sqrt{2} \pi ^2 c_w^2
   s_w^2}\\
R_2^{\bar l l h_3}&=& -\frac{e^2 G_l \left(14 s_w^2+3\right) \gamma^5}{128 \sqrt{2} \pi ^2
   c_w^2 s_w^2} \\
R_2^{\bar u d H^+}&=&  -\frac{i \delta _{i_1 i_2}}{1152   \pi ^2 c_w^2 s_w^2} 
\left(\gamma_- \left(36 c_w^2 s_w^2 \left(G_u  G_d^2+ G_d y_d y_u\right)+e^2 G_u \left(14   s_w^2+27\right)\right)\right.\nonumber\\
&&\left.+\gamma_+ \left(e^2 G_d \left(34 s_w^2-27\right)-36   c_w^2 s_w^2 \left( G_d  G_u^2+ y_d G_u y_u\right)\right)\right)\\
R_2^{\bar \nu_l l H^+}&=& \frac{i e^2 \left(14 s_w^2+3\right) G_l \gamma_+}{128 \pi ^2
   c_w^2 s_w^2} \\
R_2^{\bar d u H^-}&=&  -\frac{i \delta _{i_1 i_2} }{1152 \pi ^2   c_w^2 s_w^2}
   \left(\gamma_- \left(e^2 G_d \left(34  s_w^2-27\right)-36 c_w^2 s_w^2 \left(  G_d G_u^2+ y_d G_u y_u\right)\right)\right.\nonumber\\
   &&\left.+\gamma_+    \left(36 c_w^2 s_w^2 \left(G_u G_d^2+ G_d y_d y_u\right)+e^2 G_u \left(14 s_w^2+27\right)\right)\right)\\
R_2^{\bar l\nu_lH^-}&=& \frac{i e^2 \left(14 s_w^2+3\right) G_l \gamma_-}{128 \pi ^2
   c_w^2 s_w^2} \\
R_2^{\bar t b G^+}&=& \frac{\delta _{i_1 i_2} }{1152 \pi ^2  c_w^2 s_w^2}\left(\gamma_- \left(2 s_w^2 \left(y_t \left(-18 c_w^2
    y_b^2-7 e^2\right)-18 G_b y_b c_w^2 G_t\right)-27 e^2   y_t\right)\right.\nonumber\\
   &&\left.+\gamma_+ \left(36 c_w^2 s_w^2 \left(y_b   y_t^2+ G_b G_t y_t\right)+e^2 y_b \left(27-34 s_w^2\right)\right)\right) \\
R_2^{\bar b t G^-}&=&\frac{\delta _{i_1 i_2}}{1152 \pi ^2 c_w^2 s_w^2} \left(\gamma_- \left(y_b \left(e^2 \left(34
   s_w^2-27\right)-36 c_w^2 s_w^2 y_t^2\right)-36 G_b c_w^2 G_t s_w^2 y_t\right)\right.\nonumber\\
   &&\left.+\gamma_+ \left(y_t \left(e^2 \left(14 s_w^2+27\right)+36  s_w^2c_w^2  y_b^2\right)+36 G_b y_b G_t s_w^2
   c_w^2\right)\right),
 \eea
 where the top and bottom have been used explicitly when the vertices with the other flavors vanish but otherwise $u$, $d$, $l$ and $\nu_l$ are generic label that could be replaced by any of the tree flavors of up quark, down quark, charged lepton or neutral lepton and $y_f$ and $G_f$ are the corresponding diagonal element of the Yukawa coupling matrices.  The vertices with $h_2$ are obtained by replacing $c_1\tos s_1$ and $s_1\tos-c_1$ in the vertices with $h_1$. The same trick will be used in the following for all the vertices involving a single $h_2$ and no $h_1$.\\
 
The $R_2$ vertices for the vector interaction of the fermions are
\bea
R_2^{\bar ll A}&=&\frac{i e \left(\left(c_w^2 G_l^2 s_w^2+e^2
\left(3-2 s_w^2\right)\right)\gamma ^{\mu _3}.\gamma_- +s_w^2 \left(c_w^2 G_l^2+4 e^2\right) \gamma ^{\mu _3}.\gamma_+\right)}{32 \pi ^2 c_w^2 s_w^2}\\
R_2^{\bar\nu_l\nu_l A}&=&\frac{i e G_l^2 \gamma ^{\mu _3}.\gamma_-}{32 \pi ^2}\\
R_2^{\bar dd A}&=&  -\frac{i e \delta _{i_1 i_2}}{864 \pi ^2 c_w^2} \Bigg(\frac{\left(e^2
   \left(26 s_w^2-27\right) -9 c_w^2 s_w^2 \left(G_d^2+y_d^2-2 G_u^2-2   y_u^2\right)\right)}{s_w^2}\gamma ^{\mu _3}.\gamma_-\nonumber\\
&&   +\gamma ^{\mu _3}.\gamma_+ \left(9 c_w^2   \left(G_d^2+y_d^2\right)-4 e^2\right)\Bigg) \\
R_2^{\bar uu A}&=& \frac{i e \delta _{i_1 i_2}}{864 \pi ^2 c_w^2} 
\Bigg( \frac{\left(9 c_w^2 s_w^2 \left(G_d^2+y_d^2-2 G_u^2-2 y_u^2\right)+e^2 \left(52   s_w^2-54\right)\right)}{s_w^2}\gamma ^{\mu _3}.\gamma_-\nonumber\\
 &&-\gamma ^{\mu _3}.\gamma_+ \left(9 c_w^2   \left(G_u^2+y_u^2\right)+32 e^2\right)\Bigg)  \\
R_2^{\bar l l Z}&=& -\frac{i e \left(\gamma ^{\mu _3}.\gamma_- \left(2 c_w^2 G_l^2 s_w^4-e^2
   \left(4 c_w^4-1\right)\right)+2 s_w^4 \left(c_w^2 G_l^2+4 e^2\right) \gamma
   ^{\mu _3}.\gamma_+\right)}{64 \pi ^2 \left(c_w^2\right){}^{3/2} s_w^3}  \\
R_2^{\bar \nu_l \nu_l Z}&=& \frac{i e \gamma ^{\mu _3}.\gamma_- \left(e^2 \left(2 s_w^2-3\right)-2
   c_w^2 G_l^2 s_w^4\right)}{64 \pi ^2 \left(c_w^2\right){}^{3/2} s_w^3} \\
R_2^{\bar d d Z}&=& \frac{i e \delta _{i_1 i_2} }{1728 \pi ^2 \left(c_w^2\right){}^{3/2} s_w^3}
\Bigg(2 s_w^4\left(9 c_w^2   \left(G_d^2+y_d^2\right)-4 e^2\right) \gamma ^{\mu _3}.\gamma_+ \nonumber\\
&&+ \left(e^2 \left(52   s_w^4-132 s_w^2+81\right)-18 c_w^2 s_w^4 \left(G_d^2+y_d^2-2 G_u^2-2 y_u^2\right)\right)\gamma ^{\mu _3}.\gamma_-\Bigg) \\
R_2^{\bar uu Z}&=& \frac{i e \delta _{i_1 i_2}}{1728 \pi ^2 \left(c_w^2\right){}^{3/2} s_w^3}  
\Bigg(2 s_w^4 \left(9 c_w^2   \left(G_u^2+y_u^2\right)+32 e^2\right)\gamma ^{\mu _3}.\gamma_+\nonumber\\
&&- \left(18 c_w^2   s_w^4 \left(G_d^2+y_d^2-2 G_u^2-2 y_u^2\right)+e^2 \left(104 s_w^4-186   s_w^2+81\right)\right)\gamma ^{\mu _3}.\gamma_-
\Bigg)\\
R_2^{\bar dd G}&=& \frac{i g_s T^{{a}_3}_{i_1,i_2}}{288 \pi ^2 c_w^2} \Bigg(\frac{\left(e^2 \left(26 s_w^2-27\right)-9 c_w^2 s_w^2
   \left(G_d^2+y_d^2+G_u^2+y_u^2\right)\right)}{s_w^2}\gamma ^{\mu_3}.\gamma_- \nonumber\\
   &&-2  \left(9 c_w^2 \left(G_d^2+y_d^2\right)+2e^2\right)\gamma ^{\mu   _3}.\gamma_+\Bigg) \\
R_2^{\bar d d G}&=& \frac{i g_s T^{{a}_3}_{i_1,i_2}}{288 \pi ^2 c_w^2}
\Bigg(\frac{ \left(e^2 \left(26 s_w^2-27\right)-9 c_w^2 s_w^2   \left(G_d^2+y_d^2+G_u^2+y_u^2\right)\right)}{s_w^2}\gamma ^{\mu_3}.\gamma_-\nonumber\\
   &&-2  \left(9 c_w^2 \left(G_u^2+y_u^2\right)+8 e^2\right)\gamma ^{\mu_3}.\gamma_+\Bigg),
\eea 
where the $R_2$ vertices with the $W$-boson are omitted because there are not modified compared to the SM.
The three bosons $R_2$ vertices are also like in the SM.\\

The vector-vector-scalar $R_2$ vertices are given by
\bea  
R_2^{ AW^\pm H^\mp}&=&-\frac{i e^2 v \eta _{\mu _1,\mu _2} \left(2 G_d y_d+G_u y_u\right)}{32 \pi ^2 s_w}\\
R_2^{ Z W^\pm H^\mp}&=& \frac{i e^2 v \eta _{\mu _1,\mu _2} \left(2 G_d y_d+G_u y_u\right)}{32 \pi ^2 \sqrt{c_w^2}} \\
R_2^{A A h_1}&=& -\frac{i e^2 v \eta _{\mu _1,\mu _2} \left(-4 s_1 s_w^2 \left(G_d y_d+4 G_u y_u\right)+c_1 \left(4
   s_w^2 \left(y_d^2+4 y_u^2\right)+3 e^2\right)\right)}{96 \pi ^2 s_w^2} \\
R_2^{ A Z  h_1}&=& \frac{i e^2 v \eta _{\mu _1,\mu _2}}{192 \pi ^2   \sqrt{c_w^2} s_w^3} 
\bigg(-2 s_1 s_w^2 \left(G_d y_d \left(4 s_w^2-3\right)+2 G_u   \left(8 s_w^2-3\right) y_u\right)\nonumber\\
&&+c_1 \left(2 s_w^2 \left(y_d^2 \left(4 s_w^2-3\right)+2   \left(8 s_w^2-3\right) y_u^2\right)+e^2 \left(6 s_w^2-9\right)\right)\bigg) \\
R_2^{ Z Z h_1}&=& \frac{i e^2 v \eta _{\mu _1,\mu _2} }{96 \pi ^2 c_w^2 s_w^4} 
\bigg(s_1 s_w^2 \left(G_d y_d \left(2 s_w^2-3\right)^2+G_u \left(4 s_w^2-3\right)^2 y_u\right)\nonumber\\
&&-c_1 \left(s_w^2   \left(y_d^2 \left(2 s_w^2-3\right)^2+\left(4 s_w^2-3\right)^2y_u^2\right)+3 e^2 \left(s_w^4-3 s_w^2+2\right)\right)\bigg)\\
R_2^{W^- W^+ h_1}&=& \frac{i e^2 v \eta _{\mu _1,\mu _2} \left(3 s_1 s_w^2 \left(G_d y_d+G_u y_u\right)-c_1 \left(3
   s_w^2 \left(y_d^2+y_u^2\right)+2 e^2\right)\right)}{32 \pi ^2 s_w^4} 
\eea
while the vertices with the charged Goldstone bosons are given by their SM values.
The $R_2$ vertices for the interaction of two Goldstone bosons and a vector boson are SM-like while the other $R_2$ scalar-scalar-vector interactions are given by
\bea
R_2^{ W^\pm H^\mp G_0}&=&-\frac{e \left(\text{p}_2^{\mu _1}-\text{p}_3^{\mu _1}\right) \left(G_b y_b+G_t y_t\right)}{16 \pi   ^2 s_w}\\
R_2^{ A H^\mp G^\pm}&=& \frac{e \left(\text{p}_2^{\mu _1}-\text{p}_3^{\mu _1}\right) \left(G_b y_b+G_t y_t\right)}{8 \pi ^2} \\
R_2^{Z H^\mp G^\pm}&=&-\frac{e \left(\text{p}_2^{\mu _1}-\text{p}_3^{\mu _1}\right) \left(2 s_w^2-1\right) \left(G_b
   y_b+G_t y_t\right)}{16 \pi ^2 \sqrt{c_w^2} s_w}  \\
R_2^{A H^- H^+}&=&  \frac{i e \left(\text{p}_2^{\mu _1}-\text{p}_3^{\mu _1}\right) \left(16 c_w^2 s_w^2 \left(3
   R^d+R^l+3 R^u\right)+e^2 \left(1+12 c_w^2\right)\right)}{384 \pi ^2 c_w^2 s_w^2}\\
R_2^{ Z H^- H^+}&=& \frac{i e \left(\text{p}_2^{\mu _1}-\text{p}_3^{\mu _1}\right)}{768 \pi ^2 \left(c_w^2\right){}^{3/2} s_w^3}\big(16 s_w^2 \left(2 s_w^4-3   s_w^2+1\right) \left(3 R^d+R^l+3 R^u\right)\nonumber\\
 &&+e^2 \left(24 \left(s_w^2-2\right)   s_w^2+23\right)\big) \\
R_2^{W^\pm H^\mp h_1}&=& \frac{\mp i e \left(\text{p}_2^{\mu _1}-\text{p}_3^{\mu _1}\right)}{768 \pi ^2 c_w^2 s_w^3} \big(e^2 s_1 \left(22
   c_w^2+1\right)\nonumber\\
   &&+16 c_w^2 s_w^2 \left(s_1 \left(3 R^d+R^l+3 R^u\right)-3 c_1 \left(G_b   y_b+G_t y_t\right)\right)\big) \\
R_2^{ W^\pm H^\mp h_3}&=& 
-\frac{e \left(\text{p}_2^{\mu _1}-\text{p}_3^{\mu _1}\right) \left(16 c_w^2 s_w^2 \left(3   R^d+R^l+3 R^u\right)+e^2 \left(23-22 s_w^2\right)\right)}{768 \pi ^2 c_w^2 s_w^3}  \\ 
R_2^{ W^\pm G^\mp h_1}&=& \frac{-e \left(\text{p}_2^{\mu _1}-\text{p}_3^{\mu _1}\right)}{768 \pi ^2 c_w^2 s_w^3}  \big(c_1 \left(48 c_w^2 s_w^2   \left(y_b^2+y_t^2\right)+e^2 \left(1+22 c_w^2\right)\right)\nonumber\\&&-48 s_1 c_w^2 s_w^2 \left(G_b   y_b+G_t y_t\right)\big)\\
R_2^{ W^\pm G^\mp h_3}&=& \frac{\mp i e \left(\text{p}_2^{\mu _1}-\text{p}_3^{\mu _1}\right) \left(G_b y_b+G_t y_t\right)}{16 \pi
   ^2 s_w} \\
R_2^{Z G_0 h_1}&=&  -\frac{e \left(\text{p}_2^{\mu _1}-\text{p}_3^{\mu _1}\right)}{768 \pi ^2 \left(c_w^2\right){}^{3/2} s_w^3} \big(c_1 \left(48 c_w^2 s_w^2   \left(y_b^2+y_t^2\right)+e^2 \left(20 s_w^4-42 s_w^2+23\right)\right)\nonumber\\&&-48 s_1 c_w^2 s_w^2
   \left(G_b y_b+G_t y_t\right)\big)\\
R_2^{ AG_0 h_1}&=&  \frac{5 c_1 e^3 \left(\text{p}_3^{\mu _1}-\text{p}_2^{\mu _1}\right)}{192 \pi ^2 s_w^2}\\
R_2^{A h_1 h_3}&=&   \frac{5 e^3 s_1 \left(\text{p}_3^{\mu _1}-\text{p}_2^{\mu _1}\right)}{192 \pi ^2 s_w^2}\\
R_2^{Z h_1 h_3}&=& \frac{e \left(\text{p}_2^{\mu _1}-\text{p}_3^{\mu _1}\right)}{768 \pi ^2 \left(c_w^2\right){}^{3/2} s_w^3} \big(e^2 s_1 \left(-20 s_w^4+42   s_w^2-23\right)\nonumber\\&&-16 c_w^2 s_w^2 \left(s_1 \left(3 R^d+R^l+3 R^u\right)-3 c_1 \left(G_b   y_b+G_t y_t\right)\right)\big) , 
\eea  
where $p_k$ is the momenta of the $k^{\text{th}}$ particle and is incoming.\\

The three scalars $R_2$ vertices read
\bea
R_2^{ H^-H^+ h_1}&=& -\frac{i v}{256 \pi ^2 c_w^4 s_w^4} 
\big(2 s_1 c_w^2 s_w^2 \left(24 c_w^2 s_w^2 \left(G_d^3 y_d+G_u^3 y_u\right)+e^2   \lambda _7 \left(3-2 s_w^2\right)\right)\nonumber\\
&&+c_1 \left(-48 c_w^4 s_w^4 \left(G_d^2   \left(y_d^2+y_u^2\right)-2 G_d y_d G_u y_u+G_u^2 \left(y_d^2+y_u^2\right)\right)\right.\nonumber\\
&&\left.-2 e^2 \lambda _3 s_w^2 \left(2 s_w^4-5 s_w^2+3\right)+3 e^4 \left(6 s_w^4-8  s_w^2+3\right)\right)\big) \\
R_2^{G^\mp H^\pm h_1 }&=&  \frac{\mp v }{256 \pi ^2 c_w^2 s_w^2}
   \big(c_1 \left(48 c_w^2 s_w^2 \left(G_d y_d^3+G_u y_u^3\right)+e^2 \lambda _6   \left(6-4 s_w^2\right)\right)\nonumber\\
&&   +s_1 \left(-24 c_w^2 s_w^2 \left(G_d^2 \left(2 y_d^2-y_u^2\right)+2 G_d y_d G_u y_u-G_u^2 \left(y_d^2-2 y_u^2\right)\right)\right.\nonumber\\
&&\left.+e^2 \left(\lambda _4   \left(2 s_w^2-3\right)-8 \lambda _5 c_w^2\right)+6 e^4\right)\big)\\
R_2^{ G^- G^+ h_1}&=& \frac{i v}{256 \pi ^2 c_w^4 s_w^4}  
\big(c_1 \left(48 c_w^4 s_w^4 \left(y_d^4+y_u^4\right)+16 e^2 \lambda _1 c_w^4   s_w^2-3 e^4 \left(2 s_w^4-4 s_w^2+3\right)\right)\nonumber\\
&&-2 s_1 c_w^2 s_w^2 \left(24 c_w^2s_w^2 \left(G_d y_d^3+G_u y_u^3\right)+e^2 \lambda _6 \left(3-2   s_w^2\right)\right)\big) 
\eea
\bea
R_2^{ H^\mp G^\pm h_3}&=& -\frac{i v \left(24 c_w^2 s_w^2 \left(y_d G_u-G_d y_u\right){}^2+e^2 \left(8 \lambda _5
   c_w^2+\lambda _4 \left(2 s_w^2-3\right)\right)+6 e^4\right)}{256 \pi ^2 c_w^2 s_w^2}\\
R_2^{G_0 G_0 h_1}&=& \frac{i v}{256 \pi ^2 c_w^4 s_w^4}
 \big(c_1 \left(48 c_w^4 s_w^4 \left(y_d^4+y_u^4\right)+4 e^2 \lambda _1 c_w^2   s_w^2-3 e^4 \left(2 s_w^4-4 s_w^2+3\right)\right)\nonumber\\
 &&-2 s_1 c_w^2 s_w^2 \left(24 c_w^2 s_w^2 \left(G_d y_d^3+G_u y_u^3\right)+e^2 \lambda _6 \left(3-2 s_w^2\right)\right)\big) \\
R_2^{ h_1h_1 h_1}&=& -\frac{3 i v}{256 \pi ^2 c_w^4 s_w^4} 
\big(6 c_1^2 s_1 c_w^2 s_w^2 \left(24 c_w^2 s_w^2 \left(G_d y_d^3+G_u   y_u^3\right)+e^2 \lambda _6 \left(3-2 s_w^2\right)\right)\nonumber\\
&&+c_1 s_1^2 \left(2 e^2 c_w^2 s_w^2 \left(-3 \lambda _3-3 \lambda _4-2 \lambda _5+2 \left(\lambda _3+\lambda _4\right) s_w^2\right)\right.\nonumber\\
&&\left.-144 c_w^4 s_w^4   \left(G_d^2 y_d^2+G_u^2 y_u^2\right)+3 e^4 \left(2 s_w^4-4 s_w^2+3\right)\right)\nonumber\\
&&+2 s_1^3 c_w^2 s_w^2 \left(24 c_w^2 s_w^2 \left(G_d^3 y_d+G_u^3   y_u\right)+e^2 \lambda _7 \left(3-2 s_w^2\right)\right)\nonumber\\
&&+c_1^3 \left(-48 c_w^4 s_w^4\left(y_d^4+y_u^4\right)-4 e^2 \lambda _1 c_w^2 s_w^2+3 e^4 \left(2 s_w^4-4 s_w^2+3\right)\right)\big) \\
R_2^{ h_1 h_1 h_2}&=& \frac{-i v}{512 \pi ^2 c_w^4 s_w^4} 
\big(3 c_w^2 s_w^2s_{31} \left(e^2 \left(-2   \lambda _1+3 \lambda _3+3 \lambda _4+2 \lambda _5-2 \left(\lambda _3+\lambda _4\right)   s_w^2\right)\right.\nonumber\\
&&\left.-24 c_w^2 s_w^2 \left(-3 G_d^2 y_d^2+y_d^4-3 G_u^2 y_u^2+y_u^4\right)\right)\nonumber\\
&&+3   c_w^2 s_w^2 c_{31} \left(24 c_w^2 s_w^2 \left(G_d^3   y_d-3 G_d y_d^3-3 G_u y_u^3+G_u^3 y_u\right)+e^2 \left(3 \lambda _6-\lambda _7\right) \left(2   s_w^2-3\right)\right)\nonumber\\
&&-3 c_1 c_w^2 s_w^2 \left(24 c_w^2 s_w^2 \left(G_d^3 y_d+G_d   y_d^3+G_u y_u \left(G_u^2+y_u^2\right)\right)-e^2 \left(\lambda _6+\lambda _7\right) \left(2   s_w^2-3\right)\right)\nonumber\\
&&+s_1 \left(-72 c_w^4 s_w^4 \left(G_d^2 y_d^2+y_d^4+G_u^2   y_u^2+y_u^4\right)\right.\nonumber\\
&&\left.+e^2 c_w^2 s_w^2 \left(-6 \lambda _1-3 \lambda _3-3 \lambda _4-2 \lambda_5+2 \left(\lambda _3+\lambda _4\right) s_w^2\right)+6 e^4 \left(2 s_w^4-4   s_w^2+3\right)\right)\big)\\
R_2^{ h_1h_2 h_2}&=& \frac{-i v}{256 \pi ^2 c_w^4 s_w^4} 
\big(c_1^3 \left(3 e^4 \left(2 s_w^4-4 s_w^2+3\right)-144 c_w^4 s_w^4 \left(G_d^2 y_d^2+G_u^2 y_u^2\right)\right.\nonumber\\
&&\left.+2 e^2 c_w^2   s_w^2 \left(-3 \lambda _3-3 \lambda _4-2 \lambda _5+2 \left(\lambda _3+\lambda _4\right)   s_w^2\right)\right)\nonumber\\
&&+6 c_1^2 s_1 c_w^2 s_w^2 \left(24 c_w^2 s_w^2 \left(G_d^3 y_d-2 G_d y_d^3-2 G_u y_u^3+G_u^3 y_u\right)+e^2 \left(2  \lambda _6-\lambda _7\right) \left(2 s_w^2-3\right)\right)\nonumber\\
&&+c_1 s_1^2 \left(144 c_w^4 s_w^4   \left(2 G_d^2 y_d^2-y_d^4+2 G_u^2 y_u^2-y_u^4\right)+4 e^2 s_w^2 \left(\left(2 \lambda _5-3  \lambda _1\right) c_w^2\right.\right.\nonumber\\
&&\left.\left.+\left(\lambda _3+\lambda _4\right) \left(2 s_w^4-5  s_w^2+3\right)\right)+3 e^4 \left(2 s_w^4-4 s_w^2+3\right)\right)\nonumber\\
&&+6 s_1^3 c_w^2 s_w^2  \left(24 c_w^2 s_w^2 \left(G_d y_d^3+G_u y_u^3\right)+e^2 \lambda _6 \left(3-2   s_w^2\right)\right)\big)\\
R_2^{ h_2 h_2 h_2}&=& \frac{3 i v}{256 \pi ^2 c_w^4 s_w^4} 
\big(2 c_1^3 c_w^2 s_w^2 \left(24 c_w^2 s_w^2 \left(G_d^3 y_d+G_u^3   y_u\right)+e^2 \lambda _7 \left(3-2 s_w^2\right)\right)\nonumber\\
&&+c_1^2 s_1 \left(144 c_w^4 s_w^4   \left(G_d^2 y_d^2+G_u^2 y_u^2\right)-3 e^4 \left(2 s_w^4-4   s_w^2+3\right)\right.\nonumber\\
&&\left.+2 e^2 c_w^2 s_w^2 \left(3 \lambda _3+3 \lambda _4+2 \lambda _5-2 \left(\lambda _3+\lambda _4\right) s_w^2\right)\right)\nonumber\\
 &&+6 c_1 s_1^2 c_w^2 s_w^2 \left(24 c_w^2 s_w^2 \left(G_d y_d^3+G_u   y_u^3\right)+e^2 \lambda _6 \left(3-2 s_w^2\right)\right)\nonumber\\
 &&+s_1^3 \left(48 c_w^4 s_w^4 \left(y_d^4+y_u^4\right)+4 e^2 \lambda _1 c_w^2 s_w^2-3 e^4 \left(2 s_w^4-4s_w^2+3\right)\right)\big)\\
R_2^{h_1h_1 h_3}&=&  \frac{i v}{256 \pi ^2 c_w^4 s_w^4} 
\big(c_1 \left(48 c_w^4 s_w^4 \left(G_d^2 y_d^2+G_u^2 y_u^2\right)-3 e^4 \left(2 s_w^4-4 s_w^2+3\right)\right.\nonumber\\
&&\left.-2 e^2 c_w^2   s_w^2 \left(-3 \lambda _3-3 \lambda _4+2 \lambda _5+2 \left(\lambda _3+\lambda _4\right)s_w^2\right)\right)\nonumber\\
&&-2 s_1 c_w^2 s_w^2 \left(24  c_w^2 s_w^2 \left(G_d^3 y_d+G_u^3 y_u\right)+e^2 \lambda _7 \left(3-2  s_w^2\right)\right)\big)\\
R_2^{G_0h_1 h_3}&=& \frac{-i v}{128 \pi ^2 c_w^2 s_w^2} 
\big(2 s_1 \left(12 c_w^2 s_w^2 \left(G_d^2 y_d^2+G_u^2 y_u^2\right)+e^2 \lambda_5\right)\nonumber\\
&&+c_1 \left(e^2 \lambda _6 \left(2 s_w^2-3\right)-24 c_w^2 s_w^2 \left(G_d y_d^3+G_u y_u^3\right)\right)\big) ,
\eea
where  we remind that the vertices with a single $h_2$ and no $h_1$ are obtained by replacing $c_1\tos s_1$ and $s_1\tos-c_1$ in the vertices with a single  $h_1$ and no $h_2$.\\

Finally the $R_2$ vertices between two scalar and two vector fields are
\bea
R_2^{AW^\pm H^\mp G_0}&=&\pm\frac{e^2 \eta _{\mu _1,\mu _2} \left(3 G_d y_d+2 G_u y_u\right)}{32 \pi ^2 s_w}\\
R_2^{ZW^\pm H^\mp G_0}&=&\mp\frac{e^2 \eta _{\mu _1,\mu _2} \left(3 G_d y_d+2 G_u y_u\right)}{32 \pi ^2 \sqrt{c_w^2}}\\
R_2^{AAH^\mp G^\pm}&=&  \mp\frac{5 e^2 \eta _{\mu _1,\mu _2} \left(G_d y_d+G_u y_u\right)}{12 \pi ^2}\\
R_2^{ZAH^\mp G^\pm}&=& \pm\frac{e^2 \eta _{\mu _1,\mu _2} \left(G_d y_d \left(40 s_w^2-21\right)+2 G_u \left(20   s_w^2-9\right) y_u\right)}{96 \pi ^2 \sqrt{c_w^2} s_w} \\
R_2^{ZZH^\mp G^\pm}&=&  \mp \frac{e^2 \eta _{\mu _1,\mu _2} \left(G_d y_d \left(20 s_w^4-21 s_w^2+6\right)+2 G_u \left(10   s_w^4-9 s_w^2+3\right) y_u\right)}{48 \pi ^2 c_w^2 s_w^2}\\
R_2^{W^+W^-H^\mp G^\pm}&=&\mp\frac{e^2 \eta _{\mu _1,\mu _2} \left(G_d y_d+G_u y_u\right)}{8 \pi ^2 s_w^2}\\
R_2^{AAH^-H^+}&=& -\frac{i e^2 \eta _{\mu _1,\mu _2} \left(16 c_w^2 s_w^2 \left(5 R^d+2 R^l+5 R^u\right)+e^2   \left(22-21 s_w^2\right)\right)}{192 \pi ^2 c_w^2 s_w^2} \\
R_2^{AZH^-H^+}&=& -\frac{i e^2 \eta _{\mu _1,\mu _2} }{384 \pi ^2 \left(c_w^2\right){}^{3/2} s_w^3}\big(4 c_w^2 s_w^2 \left(\left(21-40 s_w^2\right)   R^d+\left(5-16 s_w^2\right) R^l+2 \left(9-20 s_w^2\right) R^u\right)\nonumber\\&&+e^2 \left(42 s_w^4-74   s_w^2+31\right)\big) \\
R_2^{ZZH^-H^+}&=&-\frac{i e^2 \eta _{\mu _1,\mu _2}}{768 \pi ^2 c_w^4  s_w^4}  \big(16 c_w^2 s_w^2 \left(\left(20 s_w^4-21   s_w^2+6\right) R^d+\left(8 s_w^4-5 s_w^2+2\right) R^l\right.\nonumber\\&&\left.+2 \left(10 s_w^4-9 s_w^2+3\right)   R^u\right)+e^2 \left(-84 s_w^6+208 s_w^4-162 s_w^2+39\right)\big) \\
R_2^{W^+W^-H^-H^+}&=&  -\frac{i e^2 \eta _{\mu _1,\mu _2} \left(32 c_w^2 s_w^2 \left(3 R^d+R^l+3 R^u\right)+e^2   \left(39-38 s_w^2\right)\right)}{768 \pi ^2 c_w^2 s_w^4}\\
R_2^{AW^\pm H^\mp h_1}&=&  \frac{i e^2 \eta _{\mu _1,\mu _2} \left(8 c_w^2 s_w^2 \left(s_1 \left(9 R^d+R^l+6 R^u\right)-3
   c_1 \left(3 G_b y_b+2 G_t y_t\right)\right)+e^2 s_1 \left(1+22 c_w^2\right)\right)}{768 \pi ^2
   c_w^2 s_w^3}\nonumber\\
   \\
R_2^{ZW^\pm H^\mp h_1}&=&  -\frac{i e^2 \eta _{\mu _1,\mu _2} \left(e^2 s_1 \left(22 c_w^2+1\right)+8 c_w^2 s_w^2   \left(s_1 \left(9 R^d+R^l+6 R^u\right)-3 c_1 \left(3 G_b y_b+2 G_t y_t\right)\right)\right)}{768   \pi ^2 \left(c_w^2\right){}^{3/2} s_w^2}\nonumber\\
\\
R_2^{AAG_0h_3}&=& -\frac{i e^2 \eta _{\mu _1,\mu _2} \left(G_b y_b+4 G_t y_t\right)}{24 \pi ^2} \\
R_2^{ZAG_0h_3}&=& \frac{i e^2 \eta _{\mu _1,\mu _2} \left(G_b y_b \left(4 s_w^2-3\right)+2 G_t \left(8
   s_w^2-3\right) y_t\right)}{96 \pi ^2 \sqrt{c_w^2} s_w}\\
R_2^{ZZG_0h_3}&=& -\frac{i e^2 \eta _{\mu _1,\mu _2} \left(G_b y_b \left(2 s_w^4-3 s_w^2+6\right)+2 G_t \left(4
   s_w^4-3 s_w^2+3\right) y_t\right)}{48 \pi ^2 c_w^2 s_w^2}\\
R_2^{W^+W^-G_0h_3}&=& -\frac{i e^2 \eta _{\mu _1,\mu _2} \left(G_b y_b+G_t y_t\right)}{8 \pi ^2 s_w^2}\\
R_2^{AW^\pm G^\mp h_3}&=& \frac{i e^2 \eta _{\mu _1,\mu _2} \left(3 G_b y_b+2 G_t y_t\right)}{32 \pi ^2 s_w}\\
R_2^{ZW^\pm G^\mp h_3}&=&-\frac{i e^2 \eta _{\mu _1,\mu _2} \left(3 G_b y_b+2 G_t y_t\right)}{32 \pi ^2 \sqrt{c_w^2}} \\
R_2^{AW^\pm G^\mp h_3}&=&\pm\frac{e^2 \eta _{\mu _1,\mu _2} \left(8 c_w^2 s_w^2 \left(9 R^d+R^l+6 R^u\right)+e^2 \left(23-22   s_w^2\right)\right)}{768 \pi ^2 c_w^2 s_w^3}\\
R_2^{ZW^\pm G^\mp h_3}&=& \mp\frac{e^2 \eta _{\mu _1,\mu _2} \left(8 c_w^2 s_w^2 \left(9 R^d+R^l+6 R^u\right)+e^2
   \left(23-22 s_w^2\right)\right)}{768 \pi ^2 \left(c_w^2\right){}^{3/2} s_w^2}
\eea
\bea   
R_2^{AAh_3h_3}&=& \frac{i e^2 \eta _{\mu _1,\mu _2} \left(e^2-8 s_w^2 \left(R^d+3 R^l+4 R^u\right)\right)}{192 \pi
   ^2 s_w^2}\\
R_2^{ZAh_3h_3}&=& -\frac{i e^2 \eta _{\mu _1,\mu _2} \left(2 s_w^2 \left(\left(3-4 s_w^2\right) R^d+\left(3-12
   s_w^2\right) R^l+2 \left(3-8 s_w^2\right) R^u\right)+e^2 \left(s_w^2+4\right)\right)}{192
   \pi ^2 \sqrt{c_w^2} s_w^3} \nonumber\\
   \\
R_2^{AAh_3h_3}&=& -\frac{i e^2 \eta _{\mu _1,\mu _2} }{768 \pi ^2 c_w^4 s_w^4} 
\big(16 c_w^2 s_w^2 \left(\left(2 s_w^4-3 s_w^2+6\right)   R^d+\left(6 s_w^4-3 s_w^2+2\right) R^l\right.\nonumber\\
&&\left.+2 \left(4 s_w^4-3 s_w^2+3\right) R^u\right)+e^2   \left(4 s_w^6+28 s_w^4-70 s_w^2+39\right)\big)\\
R_2^{W^+W^-h_3h_3}&=& -\frac{i e^2 \eta _{\mu _1,\mu _2} \left(32 c_w^2 s_w^2 \left(3 R^d+R^l+3 R^u\right)+e^2
   \left(39-38 s_w^2\right)\right)}{768 \pi ^2 c_w^2 s_w^4} \\
R_2^{W^\mp AG^\pm h_1}&=& \pm\frac{e^2 \eta _{\mu _1,\mu _2} \left(c_1 \left(e^2 \left(22 s_w^2-23\right)-24 c_w^2 s_w^2
   \left(3 y_b^2+2 y_t^2\right)\right)+24 s_1 c_w^2 s_w^2 \left(3 G_b y_b+2 G_t
   y_t\right)\right)}{768 \pi ^2 c_w^2 s_w^3} \nonumber\\ 
   \\
R_2^{W^\mp ZG^\pm h_1}&=& \pm\frac{e^2 \eta _{\mu _1,\mu _2} \left(c_1 \left(24 c_w^2 s_w^2 \left(3 y_b^2+2 y_t^2\right)+e^2
   \left(23-22 s_w^2\right)\right)-24 s_1 c_w^2 s_w^2 \left(3 G_b y_b+2 G_t
   y_t\right)\right)}{768 \pi ^2 \left(c_w^2\right){}^{3/2} s_w^2} \nonumber\\
   \\
R_2^{W^+W^-h_1h_1}&=&  \frac{i e^2 \eta _{\mu _1,\mu _2}}{3072 \pi ^2 c_w^2 s_w^4} \big(e^2 \left(152 s_w^2-156\right)-128 c_w^2
   s_w^2 c_1\left(s_1^2\left(3 R^d+ R^l+3 R^u\right)\right.\nonumber\\
   &&\left.+3 \left(-2 s_1 G_b y_b+c_1 y_b^2-2  s_1 G_t y_t+c_1 y_t^2\right)\right)\big)\\
R_2^{ZZh_1h_1}&=&-\frac{i e^2 \eta _{\mu _1,\mu _2}}{768 \pi ^2 c_w^4 s_w^4} 
\big(16 s_w^2 \left(c_1 y_b \left(2 s_w^4-3 s_w^2+6\right)   \left(c_1 y_b c_w^2-2 s_1 G_b c_w^2\right)\right.\nonumber\\
   &&\left.+2 s_1^2 c_w^2 \left(4 s_w^4-3 s_w^2+3\right)   R^u+2 c_1^2 c_w^2 \left(4 s_w^4-3 s_w^2+3\right) y_t^2\right.\nonumber\\
   &&\left.+2 s_{21} G_t \left(4 s_w^6-7   s_w^4+6 s_w^2-3\right) y_t\right)+16 s_1^2 c_w^2 s_w^2 \left(2 s_w^4-3 s_w^2+6\right)R^d\nonumber\\
   &&+16 s_1^2 c_w^2 s_w^2 \left(6 s_w^4-3 s_w^2+2\right) R^l+e^2 \left(4 s_w^6+28
   s_w^4-70 s_w^2+39\right)\big)\\
R_2^{ZAh_1h_1}&=&
-\frac{i e^2 \eta _{\mu _1,\mu _2} }{192 \pi ^2 \sqrt{c_w^2} s_w^3}\big(2 s_w^2 \left(2 s_1^2 \left(3-8 s_w^2\right) R^u+ s_1^2  \left(3-4  s_w^2\right) R^d+3 s_1^2  \left(1-4 s_w^2\right)   R^l\right.\nonumber\\
&&\left.+c_1 y_b \left(4 s_w^2-3\right) \left(2   s_1 G_b-c_1 y_b\right)+2 c_1 \left(3-8 s_w^2\right) y_t\left(c_1 y_t-2s_1G_t\right)\right)\nonumber\\
&&+e^2 \left(s_w^2+4\right)\big)\\
R_2^{AAh_1h_1}&=&\frac{i e^2 \eta _{\mu _1,\mu _2} \left(e^2-8 s_w^2 \left(c_1^2 y_b^2- s_{21} G_b y_b+4 c_1^2
   y_t^2-4 s_{21} G_t y_t+s_1^2 \left(R^d+3 R^l+4  R^u\right)\right)\right)}{192 \pi ^2 s_w^2}\nonumber\\
   \\
R_2^{W^+W-h_1h_2}&=&-\frac{i e^2 \eta _{\mu _1,\mu _2} \left(6 c_{21} \left(G_b y_b+G_t y_t\right)-s_{21} \left(3 R^d+R^l+3 R^u-3
   y_b^2-3 y_t^2\right)\right)}{48 \pi ^2 s_w^2}\\
R_2^{ZZh_1h_2}&=&-\frac{i e^2 \eta _{\mu _1,\mu _2} }{96 \pi ^2 c_w^2 s_w^2}
\big(2 c_{21} \left(G_b y_b \left(2 s_w^4-3 s_w^2+6\right)+2   G_t  y_t\left(4 s_w^4-3 s_w^2+3\right)\right)\nonumber\\
&&+s_{21} \left(y_b^2 \left(2  s_w^4-3s_w^2+6\right)+2 \left(4 s_w^4-3s_w^2+3\right) y_t^2+\left(6s_w^2-6-8 s_w^4\right) R^u\right.\nonumber\\
&&\left.+\left(-2 s_w^4+3 s_w^2-6\right) R^d+\left(-6 s_w^4+3 s_w^2-2\right) R^l\right)\big)\\
R_2^{ZAh_1h_2}&=&\frac{i e^2 \eta _{\mu _1,\mu _2}}{192 \pi ^2 \sqrt{c_w^2} s_w} 
\big(2 c_{21} \left(G_b y_b \left(4 s_w^2-3\right)+2 G_t   \left(8 s_w^2-3\right) y_t\right)+s_{21}\nonumber\\
&& \left(4 s_w^2 \left(y_b^2-3 R^l-4 R^u+4   y_t^2\right)+3 \left(-y_b^2+R^l+2 R^u-2 y_t^2\right)+\left(3-4 s_w^2\right)   R^d\right)\big)\\
R_2^{AAh_1h_2}&=&-\frac{i e^2 \eta _{\mu _1,\mu _2} \left(2 c_{21} \left(G_b y_b+4 G_t y_t\right)+s_{21}
   \left(y_b^2-R^d-3 R^l-4 R^u+4 y_t^2\right)\right)}{48 \pi ^2}\\
R_2^{W^+W^-h_2h_2}&=&\frac{i e^2 \eta _{\mu _1,\mu _2}}{3072 \pi ^2   c_w^2 s_w^4} 
\big(e^2 \left(152 s_w^2-156\right)\nonumber\\&&-128 c_w^2   s_w^2 \left(3 s_1 \left(2 c_1 \left(G_b y_b+G_t y_t\right)+s_1   \left(y_b^2+y_t^2\right)\right)+c_1^2\left(3  R^d+ R^l+3 R^u\right)\right)\big)\eea 
\bea
R_2^{ZZh_2h_2}&=&-\frac{i e^2 \eta _{\mu _1,\mu _2}}{768 \pi ^2 c_w^4 s_w^4} 
\big(16 s_w^2 \left(-s_1 y_b \left(2 s_w^4-3 s_w^2+6\right) \left(-2 c_1 G_b c_w^2-s_1 y_b c_w^2\right)\right.\nonumber\\
&&\left.+2 s_1^2 c_w^2 \left(4 s_w^4-3 s_w^2+3\right) y_t^2+2 s_{21} G_t \left(-4 s_w^6+7 s_w^4-6 s_w^2+3\right) y_t\right)\nonumber\\
&&+16 c_1^2 c_w^2 s_w^2 \left[\left(2   s_w^4-3 s_w^2+6\right) R^d+ \left(6 s_w^4-3 s_w^2+2\right) R^l+2\left(4 s_w^4-3 s_w^2+3\right) R^u\right]\nonumber\\
&&+e^2 \left(4 s_w^6+28 s_w^4-70 s_w^2+39\right)\big)\\
R_2^{ZAh_2h_2}&=&\frac{-i e^2 \eta _{\mu _1,\mu _2}}{192 \pi ^2 c_w s_w^3} 
\big(2 s_w^2 \left(s_1 \left(y_b \left(3-4 s_w^2\right)\left(2 c_1 G_b+s_1 y_b\right)+2 y_t\left(2 c_1 G_t+s_1y_t\right) \left(3-8 s_w^2\right)\right)\right)\nonumber\\
&&+2 c_1^2 s_w^2 \left[\left(3-4 s_w^2\right) R^d+3  \left(1-4 s_w^2\right) R^l+2  \left(3-8 s_w^2\right) R^u\right]+e^2  \left(s_w^2+4\right)\big)\\
R_2^{AAh_2h_2}&=&\frac{i e^2 \eta _{\mu _1,\mu _2} \left(e^2-8 s_w^2 \left(s_1 \left(2 c_1 \left(G_b y_b+4 G_t
   y_t\right)+s_1 \left(y_b^2+4 y_t^2\right)\right)+c_1^2 \left(R^d+3 R^l+4 R^u\right)\right)\right)}{192 \pi ^2 s_w^2},\nonumber\\
\eea
where the vertices with two Goldstone bosons have been omitted since they are identical to the SM ones.
The four scalars $R_2$ vertices have not been computed due to their size and low phenomenological relevance.
 
   
\section{2HDM UV counterterms}\label{sec:2hdmUV}

\subsection{QCD corrections}


As for the $R_2$ terms, only the UV vertices which differ from the SM are shown in this section. The first two flavors have been assumed again to be massless. Only the interactions between the quarks and the scalars have new UV divergences,
\begin{eqnarray}
UV^{\bar du H^-} &=&\frac{i g_s^2}{4\pi^2\bar\epsilon} \delta_{i_1 i_2}\left[  \left({G^d}^\dagger {V^{CKM}}^\dagger\right)_{f_1f_2}\gamma_--\left( {V^{CKM}}^\dagger G^u\right)_{f_1f_2}\gamma_+\right]\\
UV^{\bar bu H^-} &=&\frac{i g_s^2}{24\pi^2} \delta_{i_1 i_2}\left( \frac{9}{\bar\epsilon} +4 - 6 \log\left(\frac{M_b}{\mu}\right)\right)\left[ \left({G^d}^\dagger {V^{CKM}}^\dagger\right)_{3 f_2}\gamma_--\left( {V^{CKM}}^\dagger G^u\right)_{3 f_2}\gamma_+ \right]\nonumber\\
&&\\
UV^{\bar d t H^-} &=&\frac{i g_s^2}{24\pi^2} \delta_{i_1 i_2}\left( \frac{9}{\bar\epsilon} +4 - 6 \log\left(\frac{M_t}{\mu}\right)\right)\left[  \left({G^d}^\dagger {V^{CKM}}^\dagger\right)_{ f_1 3}\gamma_--\left( {V^{CKM}}^\dagger G^u\right)_{f_1 3}\gamma_+\right]\nonumber\\
&&\\
UV^{\bar b t H^-} &=&\frac{i g_s^2}{12\pi^2} \delta_{i_1 i_2}\left( \frac{6}{\bar\epsilon} +4 - 3 \log\left(\frac{M_b}{\mu}\right)- 3 \log\left(\frac{M_t}{\mu}\right)\right)\nonumber\\
&&\qquad\times\left[ \left({G^d}^\dagger {V^{CKM}}^\dagger\right)_{3,3}\gamma_- -\left( {V^{CKM}}^\dagger G^u\right)_{3,3}\gamma_+\right]\\
UV^{\bar u d H^+} &=&\frac{i g_s^2}{4\pi^2\bar\epsilon} \delta_{i_1 i_2}\left[\left( {V^{CKM}}G^d\right)_{f_1f_2}\gamma_+-\left({G^u}^\dagger V^{CKM}\right)_{f_1f_2}\gamma_-\right]\\
UV^{\bar u u h}&=&\frac{i g_s^2}{4\sqrt{2}\pi^2\bar\epsilon} \delta_{i_1 i_2}\left(G^u_{f_1f_2}\left(T_{2s_3}-iT_{3s_3}\right)\gamma_+ + {G^u}^\dagger_{f_1f_2}\left(T_{2s_3}+iT_{3s_3}\right)\gamma_-\right)\\
UV^{\bar t t h} &=&\frac{i g_s^2}{6\sqrt{2}\pi^2} \delta_{i_1 i_2}\Big[\left(y_t T_{1s_3} \gamma_ + y_t T_{1s_3} \gamma_+\right)\left( \frac{3}{\bar\epsilon} +2 - 3 \log\left(\frac{M_t}{\mu}\right)\right) \nonumber\\&&+\left(G_t\left(T_{2s_3}-iT_{3s_3}\right)\gamma_+ + {G^u}^\dagger_{3,3}\left(T_{2s_3}+iT_{3s_3}\right)\gamma_-\right)\left( \frac{3}{\bar\epsilon} +4 - 6 \log\left(\frac{M_t}{\mu}\right)\right)\Big]\\
UV^{\bar dd h} &=&\frac{i g_s^2}{4\sqrt{2}\pi^2\bar\epsilon} \delta_{i_1 i_2}\left( G^d_{f_1f_2}\left(T_{2s_3}+iT_{3s_3}\right)\gamma_++ {G^d}^\dagger_{f_1f_2}\left(T_{2s_3}-iT_{3s_3}\right)\gamma_-\right)\\
UV^{\bar b b h} &=&\frac{i g_s^2}{6\sqrt{2}\pi^2} \delta_{i_1 i_2}\Big[\left(y_b T_{1s_3} \gamma_ + y_b T_{1s_3} \gamma_+\right)\left( \frac{3}{\bar\epsilon} +2 - 3 \log\left(\frac{M_b}{\mu}\right)\right) \nonumber\\&&+\left(G_b\left(T_{2s_3}+iT_{3s_3}\right)\gamma_+ + {G^d}^\dagger_{3,3}\left(T_{2s_3}-iT_{3s_3}\right)\gamma_-\right)\left( \frac{3}{\bar\epsilon} +4 - 6 \log\left(\frac{M_b}{\mu}\right)\right)\Big],
\end{eqnarray}
where $u$ and $d$ are generic label for the massless quarks. The same factors should be applied in front of the square brackets to the vertex with the positively charged scalar as for the negatively charged scalar when one or both fermions are  massive. The same factor should also be applied to the vertex with the neutral scalar as for the negatively charged scalar when one  fermion is massive. The extra parts of these factors come from the renormalization of the wave function of the massive fermions in the on-shell scheme
\be
\delta Z_{t/b}^{L/R} = -\frac{g_s^2}{12\pi^2}\left(\frac{1}{\bar\epsilon_{UV}} +\frac{2}{\bar\epsilon_{IR}} +4 - 6 \log\left(\frac{M_{t/b}}{\mu}\right)\right).
\ee
The UV and IR poles cancel for the massless fermions
\be
\delta Z_{q}^{L/R} = -\frac{g_s^2}{12\pi^2}\left(\frac{1}{\bar\epsilon_{UV}}-\frac{1}{\bar\epsilon_{IR}} \right).
\ee
such that their wave function renormalization constants vanish. In both expression, we have add the subscript $UV$ and $IR$ to emphasize that the on-shell scheme does not only change the finite part of the counterterms but also affect the poles due to infrared divergences compared to the $\ol{MS}$. The remaining part of the poles for the quark-quark-vertices are purely  of UV origin. For the interaction between the neutral scalars and two massive quarks, the renormalization of the quark masses changes also  the renormalization of the Yukawa couplings $y_f$ since they are directly related by
\be
\delta y_{t/b} = \sqrt2\frac{\delta M_{t/b}}{v}
\ee
with
\be
\delta M_{t/b} = -\frac{g_s^2}{12\pi^2}\left(\frac{3}{\bar\epsilon_{UV}} +4 - 6 \log\left(\frac{M_{t/b}}{\mu}\right)\right).
\ee

\subsection{EW corrections}

The electroweak contributions to the UV counterterms vertices have been computed assuming CP and flavor symmetries as for the $R_2$ vertices. Furthermore, we have assumed that none of the masses are accidentally equal or vanishes and the complex mass scheme is used.
The electroweak corrections involve many more particles with different masses than the QCD corrections. As a result, the finite parts of the two-point functions probe almost all the mass hierarchies and therefore all the expressions of the $b_0$ functions. 
For example, the top mass renormalization constant gets new contributions with the charged scalar and the bottom quark in the loop as well as with all the new neutral scalars and the top quark itself,
\bea
\delta M_t\!\!&\!\!=\!\!&\!\!\delta M_t^{SM/H}+\frac{1}{64\pi^2M_t^3}\bigg[\bigg\{
\left(s_1 G_t-c_1 y_t\right)^2 \left(m_{h_1}^4 \log \left(\frac{m_{h_1}}{M_t}\right)
-\left(m_{h_1}^2-4 M_t^2\right)   \text{l}\left(M_t,m_{h_1},M_t\right)\right.\nonumber\\
&&\left.+M_t^4 \left(4 \log   \left(\frac{m_{h_1}}{M_t}\right)-4 \log \left(\frac{m_{h_1} M_t}{\mu ^2}\right)+2 \log \left(\frac{M_t}{\mu }\right)+\frac{3}{\bar\epsilon  }+7\right)
-m_{h_1}^2 M_t^2 \right.\nonumber\\
&&\left.\left(7 \log \left(\frac{m_{h_1}}{M_t}\right)-\log \left(\frac{m_{h_1} M_t}{\mu   ^2}\right)+2 \log \left(\frac{M_t}{\mu   }\right)+1\right)
\right)+h_1\tos h_2,c_1\tos s_1,s_1\tos-c_1\bigg\}\nonumber\\
&&+ G_t^2 \left(m_{h_3}^4 \log \left(\frac{m_{h_3}}{M_t}\right)-m_{h_3}^2 \text{l}\left(M_t,m_{h_3},M_t\right)-M_t^4   \left(-2 \log \left(\frac{M_t}{\mu }\right)+\frac{1}{\bar\epsilon }+1\right)\right.\nonumber\\
&&\left.-m_{h_3}^2 M_t^2 \left(-\log   \left(\frac{m_{h_3} M_t}{\mu ^2}\right)+3 \log \left(\frac{m_{h_3}}{M_t}\right)+2 \log   \left(\frac{M_t}{\mu }\right)+1\right)\right)
-4 G_b M_b G_t M_t\nonumber\\
&& \left(\text{l}\left(M_b,m_{H^+},M_t\right)+M_t^2 \left(\frac{1}{\bar\epsilon }+2-\log \left(\frac{M_b   m_{H^+}}{\mu ^2}\right)\right)+\left(M_b^2-m_{H^+}^2\right) \log   \left(\frac{m_{H^+}}{M_b}\right)\right)  \nonumber\\
&&+\left(G_b^2+G_t^2\right) \bigg(\left(-m_{H^+}^2+M_b^2+M_t^2\right)   \text{l}\left(M_b,m_{H^+},M_t\right)+M_t^4 \left(-\log \left(\frac{M_b m_{H^+}}{\mu   ^2}\right)+\frac{1}{\bar\epsilon }+2\right)\nonumber\\
&&+\left(m_{H^+}^2-M_b^2\right){}^2 \log\left(\frac{m_{H^+}}{M_b}\right)+M_t^2 M_b^2 \left(1-\log \left(\frac{M_b   m_{H^+}}{\mu ^2}\right)+\log \left(\frac{m_{H^+}}{\mu }\right)+\log \left(\frac{M_b}{\mu  }\right)\right)
\nonumber\\
  &&-M_t^2 m_{H^+}^2 \left(1-\log \left(\frac{M_b m_{H^+}}{\mu ^2}\right)+3 \log   \left(\frac{m_{H^+}}{M_b}\right)+2\log   \left(\frac{M_b}{\mu }\right)\right)\bigg)
\bigg],
\eea
where $\delta M_t^{SM/H}$ is the SM renormalization constant without the physical Higgs contribution\footnote{The Higgs contribution is obtained by replacing $\theta_1\tos 0$ and $m_{h_1}\tos m_H$ in the contribution of $h_1$} but keeping those of the Goldstone bosons, and 
\bea
\text l\left(m_1, m_2, m_3\right)&  =&  \log\left(\frac{m_1^2 + m_2^2 - m_3^2 + \sqrt{(m_1^4 + (m_2^2 - m_3^2)^2 - 2m_1^2(m_2^2 + m_3^2))}} {2 m_1 m_2}\right)\nonumber\\&&\times \sqrt{(m_1^4 + (m_2^2 - m_3^2)^2 - 2m_1^2(m_2^2 + m_3^2)).}
\eea
Only the real part of this $l$ function should be taken if the on-shell scheme was used instead since the other terms are real when the masses are real. 

Moreover, electroweak corrections mix different fields. The photon-Z mixing receives an extra contribution from the physical charged scalar 
\bea
\delta Z_{AZ}&=&\delta Z_{AZ}^{SM}-\frac{e^2 c_{2w}}{288 \pi ^2 c_w   M_Z^2 s_w}\left(3 \left(4 m_{H^+}^2-M_Z^2\right)\text{l}\left(m_{H^+},m_{H^+},M_Z\right)\right.\nonumber\\
&&\left. +24 M_Z^2 m_{H^+}^2- M_Z^4\left(\frac{3}{\bar\epsilon }-6\log \left(\frac{m_{H^+}}{\mu }\right)+8\right)\right).
\eea
Finally, the new fields two-point functions are corrected only by the electroweak interactions. For example, the wave function renormalization constant for the physical charged scalar is given by
\bea
\delta Z_{H^+H^+}\!\!&\!\!=\!\!&\!\!\frac{1}{16\pi^2}\Bigg[
-\frac{e^2c_{2w}^2}{4 c_w^2 s_w^2 m_{H^+}^4} \left(2 m_{H^+}^4 \left(\log \left(\frac{M_Z m_{H^+}}{\mu^2}\right)-\frac{1}{\bar\epsilon }\right)-m_{H^+}^2 M_Z^2\right.\nonumber\\
&&\left.+\left(m_{H^+}^2-M_Z^2\right)   \left(\text{l}\left(m_{H^+},M_Z,m_{H^+}\right)+\left(2 m_{H^+}^2-M_Z^2\right) \log
   \left(\frac{M_Z}{m_{H^+}}\right)\right)\right)\nonumber\\
&&+\Bigg\{\frac{e^2 s_1^2}{4 s_w{}^2 m_{H^+}^4 } \Bigg(\frac{ \text{l}\left(M_W,m_{h_1},m_{H^+}\right)}{ \left(\left(m_{h_1}-M_W\right)^2-m_{H^+}^2\right)
   \left(\left(m_{h_1}+M_W\right)^2-m_{H^+}^2\right)}\nonumber\\
&&\left(+M_W^2 \left(m_{h_1}^2   m_{H^+}^2-2 m_{H^+}^4+5 m_{h_1}^4\right)-2 \left(m_{h_1}^2-m_{H^+}^2\right) \left(m_{H^+}^4+m_{h_1}^4\right)+M_W^6\right.  \nonumber\\
&&\left.-M_W^4 \left(m_{H^+}^2+4 m_{h_1}^2\right)\right)+ m_{H^+}^2 \left(2m_{H^+}^2 \left(\frac{1}{\bar\epsilon }+1- \log \left(\frac{M_Wm_{h_1}}{\mu^2  }\right)\right) -2 m_{h_1}^2+M_W^2\right)\nonumber\\
&&+\log   \left(\frac{m_{h_1}}{M_W}\right) \left(M_W^4-3 m_{h_1}^2 M_W^2+2   m_{h_1}^4\right)\Bigg)+\frac{v^2 \left(\lambda _4 s_1-2 c_1 \lambda _6\right)^2}{4 m_{H^+}^4 } \nonumber\\
&&
\Bigg(\frac{M_W^4-m_{h_1}^2   \left(m_{H^+}^2+2 M_W^2\right)-M_W^2 m_{H^+}^2+m_{h_1}^4}{\left(m_{h_1}^4-2 m_{h_1}^2   \left(m_{H^+}^2+M_W^2\right)+\left(m_{H^+}^2-M_W^2\right){}^2\right)}  \text{l}\left(M_W,m_{h_1},m_{H^+}\right)\nonumber\\
&&-   \left(\left(m_{h_1}^2-M_W^2\right) \log   \left(\frac{m_{h_1}}{M_W}\right)-m_{H^+}^2\right)\Bigg)+h_1\tos h_2,c_1\tos s_1,s_1\tos-c_1 \nonumber\\
&&+h_1\tos h_3,c_1\tos 0,s_1\tos1\Bigg\}+\Bigg\{\frac{v^2 \left(c_1 \lambda _3-\lambda _7 s_1\right)^2}{m_{H^+}^4}\nonumber\\
&& \left(  \left(\left(m_{H^+}^2-m_{h_1}^2\right) \log   \left(\frac{m_{h_1}}{m_{H^+}}\right)-m_{H^+}^2\right)
-\frac{\left(m_{h_1}^2-3 m_{H^+}^2\right)}{4 m_{H^+}^2-m_{h_1}^2 }   \text{l}\left(m_{H^+},m_{h_1},m_{H^+}\right)\right)\nonumber\\
&&+h_1\tos h_2,c_1\tos s_1,s_1\tos-c_1\Bigg\}
-\sum_l G_l^2 \left(-\log \left(\frac{m_{H^+}^2}{\mu ^2}\right)+\frac{1}{\bar\epsilon }+i \pi +1\right)\nonumber\\
&&-3 \sum_{light}\left(G_d^2+G_u^2\right) \left(-\log   \left(\frac{m_{H^+}^2}{\mu ^2}\right)+\frac{1}{\bar\epsilon }+i \pi +1\right)\nonumber\\
&&-\frac{12 G_bG_tM_b M_t}{m_{H^+}^4} \left(\frac{M_b^2 \left(m_{H^+}^2+2 M_t^2\right)+M_t^2   m_{H^+}^2-M_b^4-M_t^4}{-2 M_b^2   \left(m_{H^+}^2+M_t^2\right)+\left(m_{H^+}^2-M_t^2\right){}^2+M_b^4} \text{l}\left(M_t,M_b,m_{H^+}\right)\right.\nonumber\\
&&\left.-\left(m_{H^+}^2-\left(M_b^2-M_t^2\right) \left(\log \left(\frac{M_b}{M_t}\right)\right)\right)\right)
-\frac{3 \left(G_b^2+G_t^2\right)}{m_{H^+}^4}\bigg(\text{l}\left(M_t,M_b,m_{H^+}\right)\nonumber\\
&& \frac{\left(M_b^2+M_t^2\right) \left(-M_b^2  \left(m_{H^+}^2+2 M_t^2\right)-M_t^2 m_{H^+}^2-m_{H^+}^4+M_b^4+M_t^4\right)+m_{H^+}^6 }{ -2 M_b^2   \left(m_{H^+}^2+M_t^2\right)+\left(m_{H^+}^2-M_t^2\right)^2+M_b^4} \nonumber\\ 
&&+   \left(M_b^2+M_t^2\right) \left(m_{H^+}^2-\left(M_b^2-M_t^2\right) \log   \left(\frac{M_b}{M_t}\right)\right)+m_{H^+}^4   \left(\frac{1}{\bar\epsilon   }+1-\log \left(\frac{M_bM_t}{\mu^2 }\right)\right)\bigg)
\Bigg]   \nonumber\\
\eea   
Contrary to the $R_2$ QCD vertices, only those few illustrative examples of electroweak UV counterterms are displayed due to the size of the expressions. The full list can be found on the \fr\ 2HDM web page \cite{2HDMfr}. 

\section{Conclusion}

The counterterm vertices for any model with a renormalizable Lagrangian, namely with all the operators of dimension four or less, can be obtained automatically using three packages. At this stage, the renormalizability of the model is assumed by the packages and is not checked. The model should be available in \fr\ and the Feynman gauge should be implemented. The renormalization is performed into \fr\ while the UV and $R_2$ conterterm vertices are computed in the new NLOCT package using \fa. Those vertices can be loaded afterwards into \fr\ and exported in the \ufo\ format to \amc, \gosam\ or any other one-loop tool. Consequently, NLO computations are now fully  automated for any renormalizable BSM model. \\
The code has been validated by comparing analytically the $R_2$ and UV counterterm vertices for the SM induced both by the strong and electroweak corrections to the expressions found in the literature. The $R_2$ vertices due to QCD for the MSSM have been checked as well. Finally, the SM \ufo\ with the QCD counterterms has been compared to the built in version of \amc\ and found in perfect agreement.\\
We have presented the full QCD $R_2$ and UV counterterms for the generic 2HDM. The full list of $R_2$ terms but the four scalars vertices due to the electroweak  interactions have also been given for the 2HDM with CP and flavor conservations. Only a few representative examples of the electroweak UV conterterms have been displayed due to the size of the expressions. However, the full electroweak UV counterterms have been obtained automatically using the method and the packages described in the previous sections and are publicly available on the \fr\ web site.\\
The extension of the package NLOCT to handle effective Lagrangian is in progress. This extension will additionally allow to compute the counterterm vertices in any gauge. 

\section*{Acknowledgement}

We are grateful to Adam Alloul, Neil Christensen, Federico Demartin, Claude Duhr, Rikkert Frederix, Benjamin Fuks, Thomas Hahn, Benoit Hespel, Valentin Hirschi, David Lopez, Fabio Maltoni, Olivier Mattelaer, Roberto Pittau, Hua-sheng Shao,  Eleni Vryonidou, Marco Zaro and Cen Zhang for discussion, collaboration and/or careful reading of the manuscript. This work was supported by Durham International Junior Research Fellowship and by MCnetITN FP7 Marie Curie Initial Training Network PITN-GA-2012-315877.

\bibliographystyle{ieeetr}
\bibliography{biblio}

\begin{thebibliography}{10}

\bibitem{Butterworth:2014efa}
J.~Butterworth, G.~Dissertori, S.~Dittmaier, D.~de~Florian, N.~Glover, {\em
  et~al.}, ``{Les Houches 2013: Physics at TeV Colliders: Standard Model
  Working Group Report},'' 2014.

\bibitem{Ossola:2006us}
G.~Ossola, C.~G. Papadopoulos, and R.~Pittau, ``{Reducing full one-loop
  amplitudes to scalar integrals at the integrand level},'' {\em Nucl.Phys.},
  vol.~B763, pp.~147--169, 2007.

\bibitem{Britto:2004nc}
R.~Britto, F.~Cachazo, and B.~Feng, ``{Generalized unitarity and one-loop
  amplitudes in N=4 super-Yang-Mills},'' {\em Nucl.Phys.}, vol.~B725,
  pp.~275--305, 2005.

\bibitem{Hirschi:2011pa}
V.~Hirschi, R.~Frederix, S.~Frixione, M.~V. Garzelli, F.~Maltoni, {\em et~al.},
  ``{Automation of one-loop QCD corrections},'' {\em JHEP}, vol.~1105, p.~044,
  2011.

\bibitem{Alwall:2011uj}
J.~Alwall, M.~Herquet, F.~Maltoni, O.~Mattelaer, and T.~Stelzer, ``{MadGraph 5
  : Going Beyond},'' {\em JHEP}, vol.~1106, p.~128, 2011.

\bibitem{Ossola:2008xq}
G.~Ossola, C.~G. Papadopoulos, and R.~Pittau, ``{On the Rational Terms of the
  one-loop amplitudes},'' {\em JHEP}, vol.~0805, p.~004, 2008.

\bibitem{'tHooft:1972fi}
G.~'t~Hooft and M.~Veltman, ``{Regularization and Renormalization of Gauge
  Fields},'' {\em Nucl.Phys.}, vol.~B44, pp.~189--213, 1972.

\bibitem{Kreimer:1993bh}
D.~Kreimer, ``{The Role of gamma(5) in dimensional regularization},'' 1993.

\bibitem{Korner:1991sx}
J.~Korner, D.~Kreimer, and K.~Schilcher, ``{A Practicable gamma(5) scheme in
  dimensional regularization},'' {\em Z.Phys.}, vol.~C54, pp.~503--512, 1992.

\bibitem{Kreimer:1989ke}
D.~Kreimer, ``{The $\gamma$(5) Problem and Anomalies: A Clifford Algebra
  Approach},'' {\em Phys.Lett.}, vol.~B237, p.~59, 1990.

\bibitem{Draggiotis:2009yb}
P.~Draggiotis, M.~Garzelli, C.~Papadopoulos, and R.~Pittau, ``{Feynman Rules
  for the Rational Part of the QCD 1-loop amplitudes},'' {\em JHEP}, vol.~0904,
  p.~072, 2009.

\bibitem{Garzelli:2009is}
M.~Garzelli, I.~Malamos, and R.~Pittau, ``{Feynman rules for the rational part
  of the Electroweak 1-loop amplitudes},'' {\em JHEP}, vol.~1001, p.~040, 2010.

\bibitem{Shao:2012ja}
H.-S. Shao and Y.-J. Zhang, ``{Feynman Rules for the Rational Part of One-loop
  QCD Corrections in the MSSM},'' {\em JHEP}, vol.~1206, p.~112, 2012.

\bibitem{Garzelli:2010fq}
M.~Garzelli and I.~Malamos, ``{R2SM: A Package for the analytic computation of
  the $R_2$ Rational terms in the Standard Model of the Electroweak
  interactions},'' {\em Eur.Phys.J.}, vol.~C71, p.~1605, 2011.

\bibitem{Alloul:2013bka}
A.~Alloul, N.~D. Christensen, C.~Degrande, C.~Duhr, and B.~Fuks, ``{FeynRules
  2.0 - A complete toolbox for tree-level phenomenology},'' 2013.

\bibitem{Hahn:2000kx}
T.~Hahn, ``{Generating Feynman diagrams and amplitudes with FeynArts 3},'' {\em
  Comput.Phys.Commun.}, vol.~140, pp.~418--431, 2001.

\bibitem{Degrande:2011ua}
C.~Degrande, C.~Duhr, B.~Fuks, D.~Grellscheid, O.~Mattelaer, {\em et~al.},
  ``{UFO - The Universal FeynRules Output},'' {\em Comput.Phys.Commun.},
  vol.~183, pp.~1201--1214, 2012.

\bibitem{Cullen:2011ac}
G.~Cullen, N.~Greiner, G.~Heinrich, G.~Luisoni, P.~Mastrolia, {\em et~al.},
  ``{Automated One-Loop Calculations with GoSam},'' {\em Eur.Phys.J.},
  vol.~C72, p.~1889, 2012.

\bibitem{Denner:2005fg}
A.~Denner, S.~Dittmaier, M.~Roth, and L.~Wieders, ``{Electroweak corrections to
  charged-current e+ e- to 4 fermion processes: Technical details and further
  results},'' {\em Nucl.Phys.}, vol.~B724, pp.~247--294, 2005.

\bibitem{SMfr}
\url{http://feynrules.irmp.ucl.ac.be/wiki/StandardModel}.

\bibitem{Denner:1991kt}
A.~Denner, ``{Techniques for calculation of electroweak radiative corrections
  at the one loop level and results for W physics at LEP-200},'' {\em
  Fortsch.Phys.}, vol.~41, pp.~307--420, 1993.

\bibitem{Burkhardt:1995tt}
H.~Burkhardt and B.~Pietrzyk, ``{Update of the hadronic contribution to the QED
  vacuum polarization},'' {\em Phys.Lett.}, vol.~B356, pp.~398--403, 1995.

\bibitem{Dittmaier:2001ay}
S.~Dittmaier and .~Kramer, Michael, ``{Electroweak radiative corrections to W
  boson production at hadron colliders},'' {\em Phys.Rev.}, vol.~D65,
  p.~073007, 2002.

\bibitem{Eidelman:1995ny}
S.~Eidelman and F.~Jegerlehner, ``{Hadronic contributions to g-2 of the leptons
  and to the effective fine structure constant alpha (M(z)**2)},'' {\em
  Z.Phys.}, vol.~C67, pp.~585--602, 1995.

\bibitem{Beenakker:2002nc}
W.~Beenakker, S.~Dittmaier, M.~Kramer, B.~Plumper, M.~Spira, {\em et~al.},
  ``{NLO QCD corrections to t anti-t H production in hadron collisions},'' {\em
  Nucl.Phys.}, vol.~B653, pp.~151--203, 2003.

\bibitem{Alwall:2014hca}
J.~Alwall, R.~Frederix, S.~Frixione, V.~Hirschi, F.~Maltoni, {\em et~al.},
  ``{The automated computation of tree-level and next-to-leading order
  differential cross sections, and their matching to parton shower
  simulations},'' 2014.

\bibitem{Branco:1999fs}
G.~C. Branco, L.~Lavoura, and J.~P. Silva, ``{CP Violation},'' {\em
  Int.Ser.Monogr.Phys.}, vol.~103, pp.~1--536, 1999.

\bibitem{Haber:2006ue}
H.~E. Haber and D.~O'Neil, ``{Basis-independent methods for the
  two-Higgs-doublet model. II. The Significance of tan beta},'' {\em
  Phys.Rev.}, vol.~D74, p.~015018, 2006.

\bibitem{Eriksson:2009ws}
D.~Eriksson, J.~Rathsman, and O.~Stal, ``{2HDMC: Two-Higgs-Doublet Model
  Calculator Physics and Manual},'' {\em Comput.Phys.Commun.}, vol.~181,
  pp.~189--205, 2010.

\bibitem{Plehn:2002vy}
T.~Plehn, ``{Charged Higgs boson production in bottom gluon fusion},'' {\em
  Phys.Rev.}, vol.~D67, p.~014018, 2003.

\bibitem{Berger:2003sm}
E.~L. Berger, T.~Han, J.~Jiang, and T.~Plehn, ``{Associated production of a top
  quark and a charged Higgs boson},'' {\em Phys.Rev.}, vol.~D71, p.~115012,
  2005.

\bibitem{Gerard:2007kn}
J.-M. Gerard and M.~Herquet, ``{A Twisted custodial symmetry in the
  two-Higgs-doublet model},'' {\em Phys.Rev.Lett.}, vol.~98, p.~251802, 2007.

\bibitem{2HDMfr}
\url{http://feynrules.irmp.ucl.ac.be/wiki/2HDM}.

\end{thebibliography}

\end{document}